\documentclass[]{imsart}

\RequirePackage{amsthm,amsmath,amsfonts,amssymb}
\RequirePackage[authoryear]{natbib}%% uncomment this for author-year citations
\RequirePackage[colorlinks,citecolor=blue,urlcolor=blue,backref=page,backref=page]{hyperref}
\RequirePackage{graphicx}
\usepackage{mathtools,prodint}

\startlocaldefs
\theoremstyle{plain}
\newtheorem{axiom}{Axiom}

\newtheorem{remark}[axiom]{Remark}
\newtheorem{theorem}{Theorem}[section]
\newtheorem{lemma}[theorem]{Lemma}
\newtheorem{corollary}[theorem]{Corollary}
\theoremstyle{definition}
\newtheorem{definition}[theorem]{Definition}

\theoremstyle{remark}

\newcommand{\pr}{{\mathbb{P}}}
\newcommand{\E}{{\mathbb{E}}}

\newcommand{\dd}{{\mathrm{d}}}
\newcommand{\ov}[1]{\overline{#1}}
\newcommand{\R}{\mathbb{R}}

\newcommand{\Be}{\mathrm{Be}}
\newcommand{\eqd}{\overset{d}{=}}
\newcommand{\ve}{\varepsilon}

\endlocaldefs

\begin{document}

\begin{frontmatter}
\title{Bayesian non-parametric survival estimation: stochastic hyperparameter sequences and distribution splicing}
%\title{A sample article title with some additional note\thanksref{t1}}
\runtitle{Bayesian non-parametric splicing}
%\thankstext{T1}{A sample additional note to the title.}

\begin{aug}
%%%%%%%%%%%%%%%%%%%%%%%%%%%%%%%%%%%%%%%%%%%%%%%
%% Only one address is permitted per author. %%
%% Only division, organization and e-mail is %%
%% included in the address.                  %%
%% Additional information can be included in %%
\author[A]{\fnms{Martin}~\snm{Bladt}\ead[label=e1]{martinbladt@math.ku.dk}}
\and
\author[B]{\fnms{Jorge}~\snm{Gonz\'alez C\'azares}\ead[label=e2]{jorge.gonzalez@sigma.iimas.unam.mx}}
%%%%%%%%%%%%%%%%%%%%%%%%%%%%%%%%%%%%%%%%%%%%%%
%% Addresses                                %%
%%%%%%%%%%%%%%%%%%%%%%%%%%%%%%%%%%%%%%%%%%%%%%
\address[A]{Department of Mathematical Sciences, University of Copenhagen\printead[presep={,\ }]{e1}}

\address[B]{Department of Probability and Statistics, IIMAS, UNAM\printead[presep={,\ }]{e2}}
\runauthor{M. Bladt and J. Gonz\'alez C\'azares}
\end{aug}

\begin{abstract}
A Bayesian non-parametric framework for studying time-to-event data is proposed, where the prior distribution is allowed to depend on an additional random source, and may update with the sample size. Such scenarios are natural, for instance, when considering empirical Bayes techniques or dynamic expert information. In this context, a natural stochastic class for studying the cumulative hazard function are conditionally inhomogeneous independent increment processes with non-decreasing sample paths, also known as mixed time-inhomogeneous subordinators or mixed non-decreasing additive processes. 

The asymptotic behaviour is studied by showing that Bayesian consistency and Bernstein--von~Mises theorems may be recovered under suitable conditions on the asymptotic negligibility of the stochastic prior sequences. The non-asymptotic behaviour of the posterior is also considered. Namely, upon conditioning, an efficient and exact simulation algorithm for the paths of the Beta L\'evy process is provided. As a natural application, it is shown how the model can provide an appropriate definition of non-parametric spliced models. Spliced models target data where an accurate global description of both the body and tail of the distribution is desirable. The Bayesian non-parametric nature of the proposed estimators can offer conceptual and numerical alternatives to their parametric counterparts.
\end{abstract}

\begin{keyword}[class=MSC]
\kwd[Primary: ]{62F15}
\kwd{62G05}
\kwd{62N02}
\kwd[; secondary: ]{00A72}
\kwd{62E20}
\end{keyword}

\begin{keyword}
\kwd{Bayesian non-parametric splicing} 
\kwd{Bernstein--von Mises} \kwd{exact simulation} 
\kwd{Beta-Stacy process} \kwd{hybrid Bayesian models}
\end{keyword}
\end{frontmatter}

\section{Introduction}

This paper explores survival models within the framework of Bayesian non-parametric statistics, known as neutral to the right models~(\citealp{MR373081,MR1714716}), where the prior itself is stochastic and may even be estimated from the same data, other datasets, or external prior information. Such settings often arise in domains where the traditional Bayesian paradigm -- relying purely on expert-driven priors -- is impractical or unrealistic and where the standard tools of objective Bayesian analysis (i.e., the construction of noninformative priors,~\citealp{MR2221271,MR3807861}) cannot be implemented due to the infinite dimensionality of the prior space. Specifically, we consider a scenario in which the prior evolves dynamically with the number of data points, and its parameters are treated as stochastic. 

Our approach stands in contrast with the usual Bayesian philosophy wherein the prior is not dependent on the data. Instead, the motivation for this hybrid scheme is purely pragmatic: we incorporate the strengths of the usual Bayesian non-parametric methods in the bulk of the domain of the data (which, in many ways, is comparable to the frequentist counterparts such as the Kaplan--Meier or Nelson--Aalen estimators,~\citealp{MR2089131,BvM-propHazard,DEBLASI20092316}) and, for the tail behaviour, we introduce a different learning mechanism that informs the choice of baseline distribution (which would correspond to the prior in the usual models), thus forcing the posterior mean to have the desired form. Contrary to hierarchical models~(\citealp{MR4255111}) that still fall within the classical Bayesian philosophy, our approach is closer to frequentist distribution splicing, where the imputed portion of the distribution is the tail and depends on an estimated parameter. More broadly, similar constructions can be applied in any context where a particular aspect of the survival curve is of primary interest to the modeler and can be forced into the posterior by appropriately choosing the baseline distribution. 

One challenge in the widespread adoption of Bayesian statistics in industry is demonstrating its reliability compared to established frequentist methods as well as the demanding computational cost of its fitting. The former has been widely addressed in the more classical nonparametric Bayesian contexts (even in the multidimensional or mixture cases, see~\citealp{MR1701105,MR1986664,MR2399197,MR2533475,MR4480729,rivapalacio2024}), which play a crucial role in building this confidence by showing that Bayesian models exhibit similar large sample behavior to their frequentist counterparts. Here, we pursue the same asymptotic results, such as consistency and the Bernstein–von Mises theorem for both the cumulative hazard and survival functions, giving sufficient conditions for the consistency and asymptotic normality in terms of growth conditions of the baseline model (see Theorems~\ref{thm:consist} and~\ref{thm:BvM1} below). In particular, they ensure that Bayesian credible regions asymptotically align with traditional confidence intervals, while providing more adaptable methods in finite-samples, particularly when extrapolating to regions outside of the range of the observed data. (See, for instance, Figures~\ref{fig:simulated-pareto},~\ref{fig:simulated-weibull} and~\ref{fig:diabetic} below.)

We specialise our analysis for the popular Beta-Stacy model (i.e., Beta L\'evy process, see, e.g.~\citealp{MR1463574}), and develop exact efficient simulation algorithms for (conditional) Beta L\'evy processes, enabling efficient sampling from both prior and posterior distributions of the hazard. Similar algorithms are derived for the associated survival function process. This represents a technical yet important advancement over existing approximate simulation methods, which often involve complex or intractable errors (see~\citealp{MR4148234} for some efforts towards controlling the Wasserstein error of certain approximations), making error control difficult or costly. Eliminating the sampling error yields more reliable Monte Carlo estimators (see, e.g.,~\citealp[\S3.3.1]{GCFirstPassage} and~\citealp{Dai19}) but can be difficult or computationally demanding. The algorithm presented here imposes the prior tuning function to be piecewise constant (which is then also the case for the posterior tuning function), but is efficient and exact. Moreover, the algorithm can be easily extended to tuning functions that are piecewise linear or piecewise defined with elementary forms in each piece.

Spliced distributions can be used in medical sciences for analyzing patient survival data~(\citealp{MR3910216}), where early-stage and late-stage survival times often follow different statistical behaviors. Patients diagnosed at an early stage may have a relatively high survival probability, best modeled with a light-tailed distribution or with a non-parametric model, while those in advanced stages face a different rate of decline, requiring a tail-specific model to capture long-term risk. Similarly, in insurance and actuarial science, they are essential for modeling claim severities, where frequent, low-cost claims follow one pattern while rare, high-cost claims exhibit usually heavy tails, see, e.g.~\cite{Gay_2005,reynkens2017modelling}. Models in the literature are often frequentist and parametric. To the best of our knowledge, we provide in this paper the fist Bayesian non-parametric definition of a spliced model.

Our work is related to Bayesian hierarchical models in the sense that the hyperparameters are considered as random variables in both approaches, see e.g.~\cite{teh2010hierarchical} for such a hierarchical model within the Bayesian non-parametric framework. The key feature which distinguishes the two approaches is that hierarchical frameworks use fixed and nested priors, while we propose unspecified stochastic (and dynamic) priors influenced by external sources. In particular, we envision prior adaptation as more data arrives or as conditions change, for instance by combining expert information and censored data, as was done in~\cite{bladt2020combined} in a parametric context. Furthermore, we have specialised in the estimation of cumulative hazard processes and survival functions.

The paper is organised as follows. We begin with providing the model specification in Section~\ref{sec:2}, as well as introducing essential concepts in survival analysis, L\'evy process priors, posterior distributions, and conditional Beta L\'evy processes.
In Section~\ref{sec:3a}, we study Bayesian consistency and in Section~\ref{sec:3b} we present the Bernstein--von~Mises theorem, linking the asymptotic behavior of Bayesian estimators with frequentist approaches. Exact simulation methods for survival probabilities and hazard functions are constructed in Section~\ref{sec:4}. This is followed by the specification of spliced models in Section~\ref{sec:5}. Numerical studies and a real data analysis are presented in Section~\ref{sec:6}. Finally, Section~\ref{sec:8} concludes. Proofs may be found in the Supplement.

\section{The Model Specification}\label{sec:2}

This section introduces a Bayesian model with random hyperparameters for estimation of time-to-event data. In particular, all results also hold when no censoring is present. Most background material can be found in  \cite{AndersenBorganGillKeiding1993}, \cite{Vaart1998} and \cite{ghosal2017fundamentals} (which itself heavily relies on~\citealp{hjort1990nonparametric,kim1999nonparametric,MR1463574}) and \cite{KimLee2004} for the case of deterministic hyperparameters (e.g., baseline hazard) not changing with the sample size.

\subsection{Preliminaries and notation in survival analysis}

Let $(\Omega,\mathcal{F},\pr)$ be a probability space, and let $X$ be a random variable of interest, with cumulative distribution function (cdf) $F(x)=\pr(X\le x)$. For simplicity, and without loss of generality, assume that the support of $X$ is given by $\text{supp}(X)=(0,\infty)$. We are interested in the non-parametric Bayesian estimation of $F$, or equivalently of that of the survival function $\ov{F}=1-F$, when observations are possibly right censored. 

Here, we make the \emph{independent right censoring} assumption, which entails the existence of a censoring random variable $C$ with $\text{supp}(C)=(0,\infty]$ such that we only observe the tuple
\begin{align*}
T=\min\{X,C\},\quad \delta=I(X=T).
\end{align*}
When $\text{supp}(C)=\{\infty\}$ we retrieve the \emph{fully-observed} case. 

Conceptually and mathematically, it is convenient to tackle the estimation of $F$ indirectly, through the cumulative hazard function, given by 
\begin{align}\label{integral}
H(t)=\int_{(0,t]}\frac{1}{\ov{F}(s-)}\dd F(s).
\end{align}
One can then return to the original scale through the inverse operation known as the \emph{product integral}, commonly denoted by the $\prodi$ symbol, such that
\begin{align}\label{prod_integral}
\ov{F}(t)
=\prodi_{(0,t]}(1-\dd H(s))
\coloneqq\exp(-H^c(t))\prod_{s\in(0,t]}(1-\Delta H(s)),
\end{align}
where $\Delta H(s) = H(s)-H(s-)$ denotes the jump of the process $H$ at time $s$, and $H^c(t)=H(t)-\sum_{s\in(0,t]}\Delta H(s)$ corresponds to the \emph{continuous part of} $H$. Note here that the jumps of $H$ are bounded by $1$ (otherwise $F$ would be decreasing at that jump time) and, if a jump of size $1$ occurs, the function $F$ becomes identically $1$ thereafter.

The functionals in \eqref{integral} and \eqref{prod_integral} have favourable properties which allow studying the cumulative hazard or the cdf rather interchangeably. For instance, Hadamard differentiability holds for integrals of the form \eqref{integral}, cf. Lemma 20.10 in \cite{Vaart1998}, as well as for the product integral in \eqref{prod_integral}, cf. \cite{GillJohansen1990}.

Assume we have an independent and identically distributed (i.i.d.) sample of size $n\in\mathbb{N}$:
\begin{align*}
\mathcal{D}_n=\{(T_i,\delta_i),\, i=1,\dots,n\}.
\end{align*}
The standard non-parametric estimator for $F$ is given by the Kaplan--Meier estimator $F_n$, which under the random censoring assumption can be derived as the \emph{non-parametric maximum likelihood estimate} of the likelihood function
\begin{align*}
\ell(F)=\prod_{i=1}^n f(X_i)^{\delta_i} \ov{F}(C_i)^{1-\delta_i}.
\end{align*}
The derivation of this type of likelihood functions can be found, for instance, in \cite{AndersenBorganGillKeiding1993}. Define the counting processes
\begin{align*}
N_n(t)=\sum_{i=1}^n I(T_i\le t)\delta_i, \quad Y_n(t)=\sum_{i=1}^n I(T_i\ge t),
\end{align*}
which denote the number of events before time $t$, and the number of individuals \emph{at risk} of having an event at time $t$, respectively. Then, the solution is given by
\begin{align*}
\ov{F}_n(t)=\prodi_{(0,t]}(1-\dd H_n(s)), 
\quad \mbox{where}\quad 
H_n=\int_{(0,t]}\frac{1}{Y_n(s)}\dd N_n(s),
\end{align*}
which is unique until the largest $T_i$ in the sample, whereafter it is undefined whenever the status indicator corresponding to the largest event time $T_i$ equals zero $\delta_i=0$, i.e. if the largest observation is right-censored. The estimator $H_n$ is known as the Nelson--Aalen estimator.

\subsection{Conditional L\'evy process survival priors}

In the Bayesian non-parametric setup, we assume $H$ is specified through a non-decreasing stochastic process. Since computing the posterior distribution for an arbitrary prior distribution on $H$ is difficult, we consider the class of conditionally right-continuous with left limits stochastic processes with independent increments, i.e. conditional L\'evy processes. In some cases, this class enjoys conjugacy, simplifying the Bayesian analysis and allowing for asymptotic analysis of the posterior for growing sample size.

Let $(\nu_n^H)$ be a collection of locally finite random measures on $D=\R_+\times (0,1]$ (denoting $\R_+\coloneqq (0,\infty)$) with associated laws denoted by $\mathcal{L}_n(B)=\pr(\nu_n^H\in B)$, for any measurable set $B$ in the space $M$ of measures on $D$ endowed with the $\sigma$-algebra $\mathcal{M}$ induced by the evaluation maps $\nu\mapsto \nu(A)$, with $\nu\in M$ and $A$ a Borel subset of $D$. For fixed $n$ and assuming $\int_{(0,t)\times\R_+}x\,\nu_n^H(\dd s,\dd x)<\infty$ for all $t>0$ a.s., we specify $H|\nu_n^H$ as a conditional L\'evy process with L\'evy measure $\nu_n^H$. The non-decreasing constraint on the process allows the following conditional \emph{L\'evy--It\^o representation} (see, e.g.~\citealp[Thm~16.2]{MR4226142}):
\begin{align}\label{pois_rep}
H(t)|\nu_n^H=\int_{(0,t]}\int x M_n (\dd s, \dd x),
\qquad M_n=M_n^c+M_n^d
\end{align}
where, conditional on $\nu_n^H$, $M_n^c$ is a Poisson random measure on $D$ (in other words, a Cox process directed by $\nu_n^H$), and 
\begin{align*}
M_n^d=\sum_j \delta_{(t^n_j,\Delta H(t^n_j))},
\end{align*}
where $t^n_j$ are the jump times of $H$. Though it might seem artificial to include atoms in a prior distribution, we include them since they naturally arise for the posterior distribution. The fixed jump random variables $\Delta H(t^n_j)$ can be taken as arbitrary conditionally independent non-negative random variables, which are conditionally independent of $M_n^c$. 

The conditional mean measures, or intensities, are random measures themselves, satisfying $\nu_n^H=\E [M_n|\nu_n^H]$, $\nu_n^{H,c}=\E [M_n^c|\nu_n^H]$, $\nu_n^{H,d}=\E  [M_n^d|\nu_n^H]$ as well as $\nu_n^H =\nu_n^{H,c} +\nu_n^{H,d} $. The supports of $\nu_n^H$, $\nu_n^{H,c}$ and $\nu_n^{H,d}$ are random sets with
\begin{align*}
\mbox{supp}(\nu_n^{H,d})\subset \bigcup_j\{t^n_j\}\times(0,1],
\qquad \mbox{supp}(\nu_n^H),\,\mbox{supp}(\nu_n^{H,c})\subset D.
\end{align*}

Then, using Lemma 3.1 in \cite{kallenberg2017random}, we may write the conditional \emph{L\'evy--Khintchine} formula $\mathcal{L}_n-\mbox{a.s.}$ as
\begin{align*}
&-\log \E[\exp(-\theta H(t))|  \nu_{n}^{H}]\\
&=\int_{(0,t]\times\R_+} (1-\exp(-\theta x))\nu_n^{H,c}(\dd s, \dd x)-\sum_{s\le t} \int_{\R_+} \exp(-\theta x) \nu_n^{H,d}( \{s\}, \dd x),
\quad \theta,t\ge 0.
\end{align*}
The conditional mean and variance of these processes are instrumental when considering posterior distributions since they allow for the study of consistency. Roughly speaking, if the mean is centered and the variance vanishes, the posterior distribution degenerates to the true hazard function $H$. Their computation can be done via differentiation at $\theta=0$, and is $\mathcal{L}_n-\mbox{a.s.}$ given by
\begin{align}
\E_{\nu_n^H} H(t)&=\int_{(0,t]\times\R_+}x\,\nu_n^{H}(\dd s, \dd x),\label{eq:mean}\\
\mathrm{var}_{\nu_n^H}(H(t))&=\int_{(0,t]\times\R_+} x^2\,\nu_n^{H}(\dd s, \dd x)-\sum_{s\le t} \bigg(\int_{\R_+} x\,\nu_n^{H,d}(\{s\}, \dd x)\bigg)^2.\label{eq:var}
\end{align}
Here and in the sequel we use the notation $\E_{\nu_n^H}[\:\cdot\:]=\E[\:\cdot\:|\nu_n^H]$, and similarly for $\mathrm{var}_{\nu_n^H}$ and $\pr_{\nu_n^H}$. Observe that the above formulas also provide insight as to how the measure $\nu_n^H$ is acting on the prior distribution.

The following result is a slight adaptation of Lemma 13.9 in \cite{ghosal2017fundamentals}, and gives a sufficient condition on a general measure $\nu_n^H$ with atomless and atomic components given by $\nu_n^{H,c}$ and $\nu_n^{H,d}$, to be a conditional intensity measures, and to moreover be a valid cumulative hazard leading to a proper cdf.

\begin{lemma}[Proper distribution]
Let $\nu_n^H$ satisfy for each $t>0$
\begin{align*}
\int_{(0,t]\times(0,1]} x\,\nu_n^{H,c}(\dd s, \dd x)<\infty, 
\quad \mbox{and} \quad 
\nu_n^{H,d}(\{t\}\times(0,1])\le 1,\quad \pr_{\mathcal{L}_n}-\mbox{a.s.}
\end{align*} 
Then $\nu_n^H$ is the conditional L\'evy measure of a L\'evy process. If moreover 
\begin{align*}
\mathrm{supp}(\nu_n^H)\subset \R_+\times(0,1], \quad \mbox{and} \quad \int_{\R_+\times(0,1]} x\,\nu_n^H(\dd s, \dd x)=\infty, \quad \pr_{\mathcal{L}_n}-\mbox{a.s.}
\end{align*} 
then the function $F(t)= 1-\prodi_{(0,t]}(1-\dd H(s))$ is non-negative, right-continuous, non-decreasing and $F(t)\to 1$, as $t\to\infty$, $\pr_{\mathcal{L}_n}$-a.s.
\end{lemma}

\subsection{Posterior distribution}
Likewise, the conjugacy result of \cite{hjort1990nonparametric} (see also \citealp[Thm~13.15]{ghosal2017fundamentals}) transfers directly to our setting, which provides the main motivation to use conditional L\'evy processes as a prior family of stochastic processes.

For this purpose, denote by $\tilde{\mathcal{L}}_n$ the joint law of $\nu_n^H$ and the elements in $\mathcal{D}_n$.

\begin{theorem}[Conjugacy]\label{thm:conjugacy}
Let $H|\nu_n^H$ be conditionally L\'evy with intensity random measure
\begin{align*}
\nu_n^H(\dd t,\dd x)=\rho_n(\dd x|t)\Lambda_n(\dd t)
\end{align*}
such that, $\pr_{\mathcal{L}_n}$-a.s., the function of measures $t\mapsto x\,\rho_n(\dd x|t)$ is weakly continuous\footnote{We say that a function $t\mapsto f(t)$ is weakly continuous if its set of discontinuities is finite.} and $\Lambda_n$ has no atoms. Then the posterior distribution process is conditional L\'evy with intensity $\tilde{\nu}_n^{H}$ measure satisfying, $\tilde{\mathcal{L}}_n-\mbox{a.s.}$,
\begin{align*}
\tilde{\nu}_n^{H, c}(\dd t,\dd x)
&=(1-s)^{Y_n(t)}\rho_n(\dd x|t)\Lambda_n(\dd t),\\
\tilde{\nu}_n^{H, d}(\{t\},\dd x)
&\,\,\propto\,\, x^{\Delta N_n(t)}(1-x)^{Y_n(t)-\Delta N_n(t)}\rho_n(\dd x|t).
\end{align*}
\end{theorem}

\subsection{Conditional Beta L\'evy processes}
A rather popular class of processes in survival Bayesian non-parametrics is that of Beta L\'evy processes, which is conveniently parametrized by two functions. The one function regulates, roughly speaking the magnitude of the posterior variance, while the other regulates the posterior mean. Here, we allow these parameters to be random and to depend on the sample size $n$, which is helpful when defining spliced models.
\begin{definition}[Conditional Beta process]
A conditional Beta process with parameters $(c_n,\Lambda_n)$ has measures given by
\begin{align*}
\nu_n^{H,c}(\dd t, \dd x) 
&= c_n(t)(1-x)^{c_n(t)-1} \frac{\dd x}{x} \dd\Lambda_n(t),\\
\nu_n^{H,d}\big(\{t\}, \cdot\big)
&= \mathrm{Be}(c_n(t)\Delta \Lambda_n(t),c_n(t)(1-\Delta \Lambda_n(t))).
\end{align*}
Then we have that
\begin{equation}
\label{eq:posterior_mean_var}
\E_{\nu_n^H} H(t)=\Lambda_n(t), \quad\mathrm{var}_{\nu_n^H}(H(t))=\int_{(0,t]}\frac{1-\Delta\Lambda_n}{c_n+1}\dd \Lambda_n
\end{equation}
\end{definition}

It follows that for $c_n$ and $\Lambda_n$ satisfying the conditions of Theorem~\ref{thm:conjugacy}, the posterior distribution process is again conditional Beta with new parameters
\begin{equation}\label{posterior_beta}
\tilde c_n(t)= c_n(t)+Y_n(t),\quad \tilde\Lambda_n(t)=\int_{(0,t]}\frac{c_n\dd \Lambda_n+\dd N_n}{c_n+Y_n}.
\end{equation}
The posterior mean and variance in themselves are quantities that depend on two sources of randomness (though no longer the source coming from the stochastic processes): the data and the expert information encoded in the parameters $(c_n,\Lambda_n)$. Thus, the posterior mean, for instance, is defined to sensibly include dynamic expert information into the Nelson--Aalen cumulative hazard estimator.

\begin{figure}[htb!]
    \centering{}
    \includegraphics[width=5cm,trim={0.5cm 5.5cm 5cm 11cm},clip]{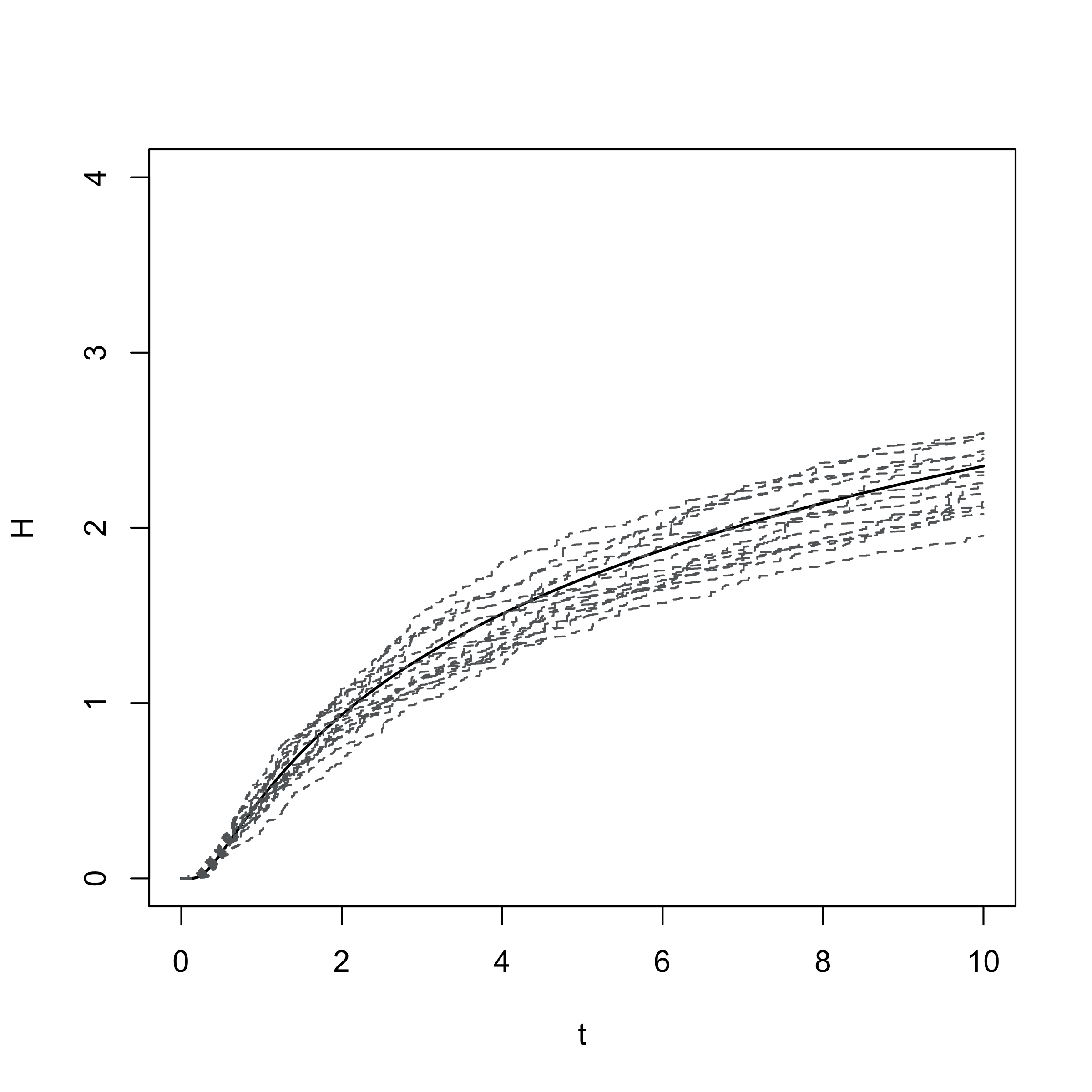}
    \includegraphics[width=5cm,trim={0.5cm 5.5cm 5cm 11cm},clip]{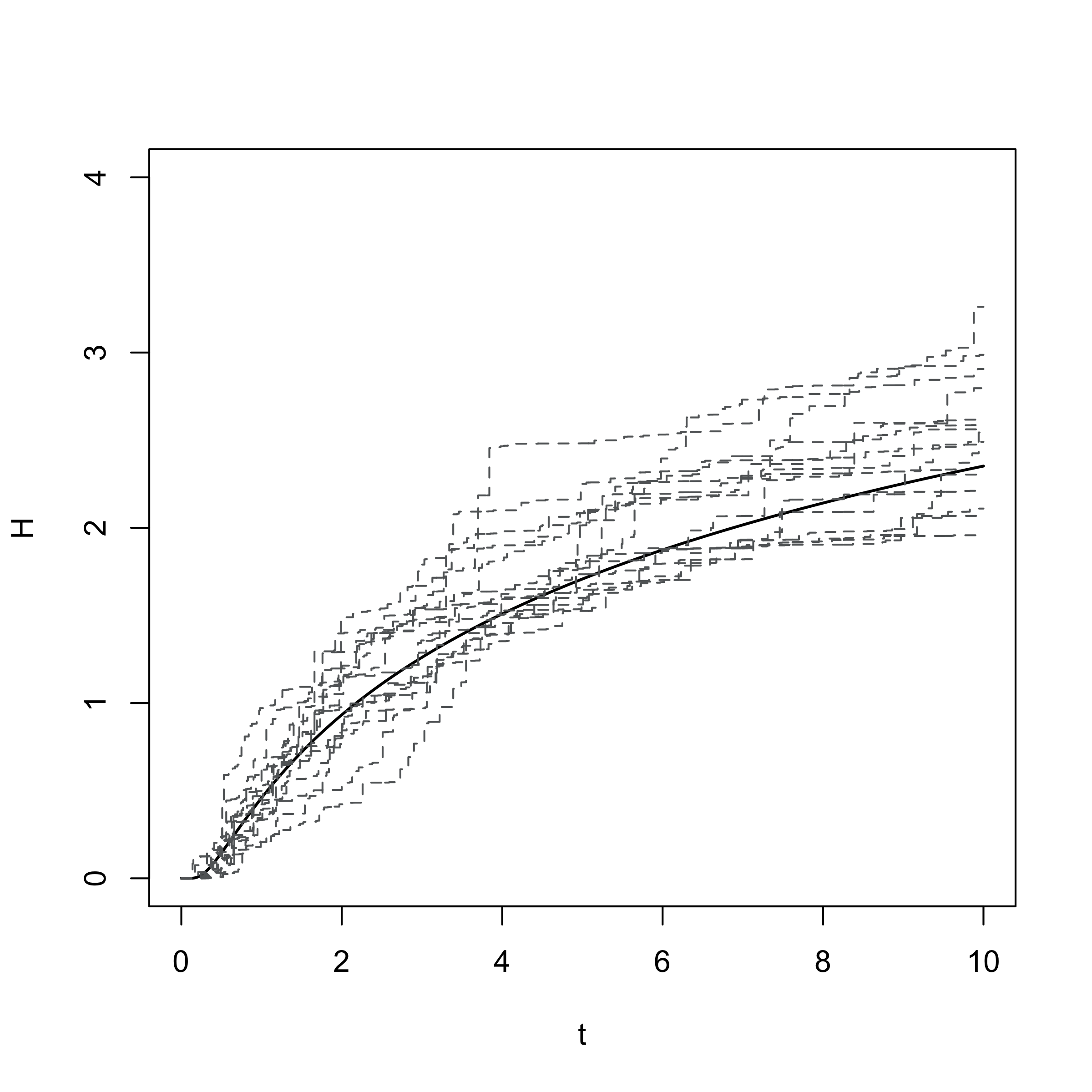}
    \includegraphics[width=5cm,trim={0.5cm 3.5cm 5cm 11cm},clip]{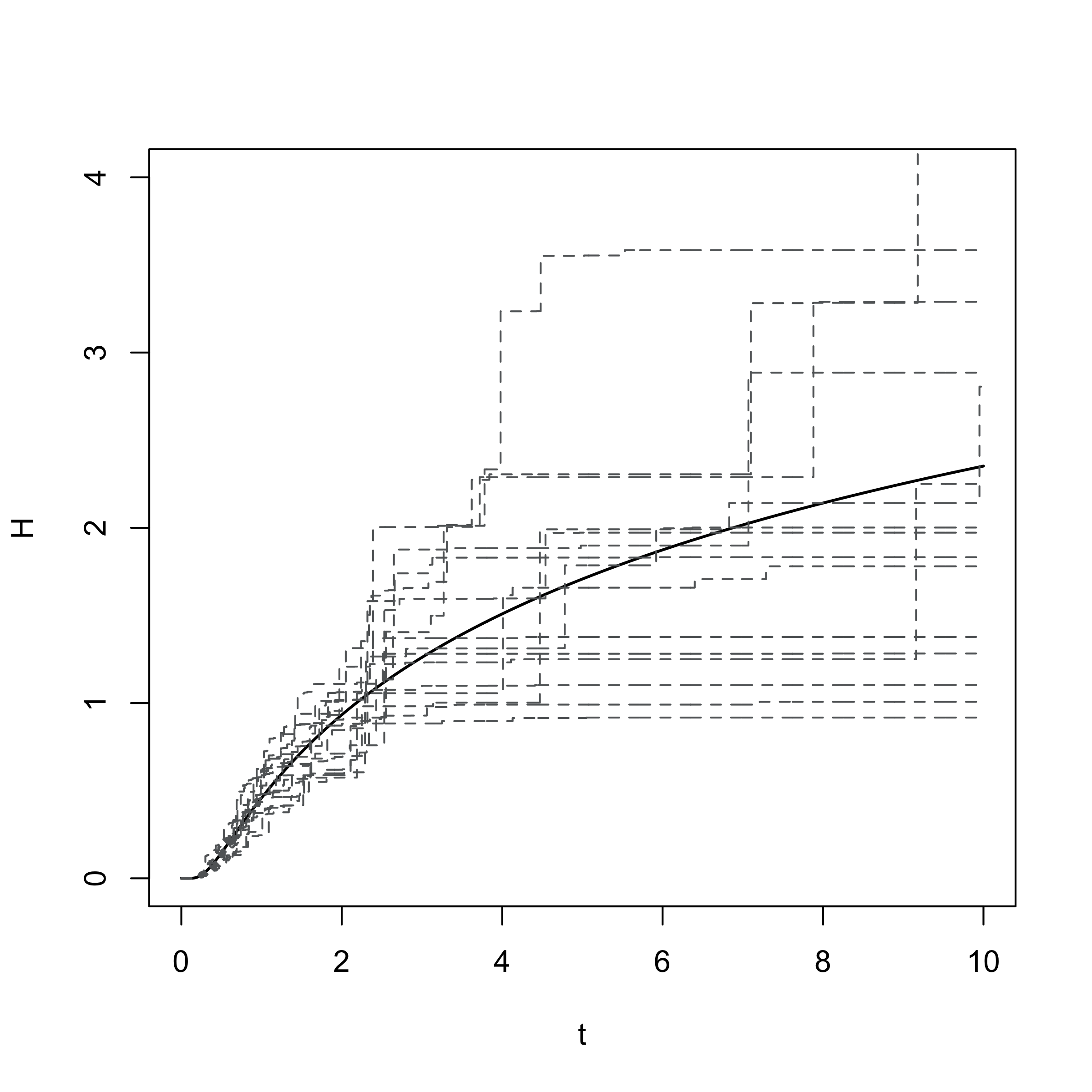}
    \includegraphics[width=5cm,trim={0.5cm 3.5cm 5cm 11cm},clip]{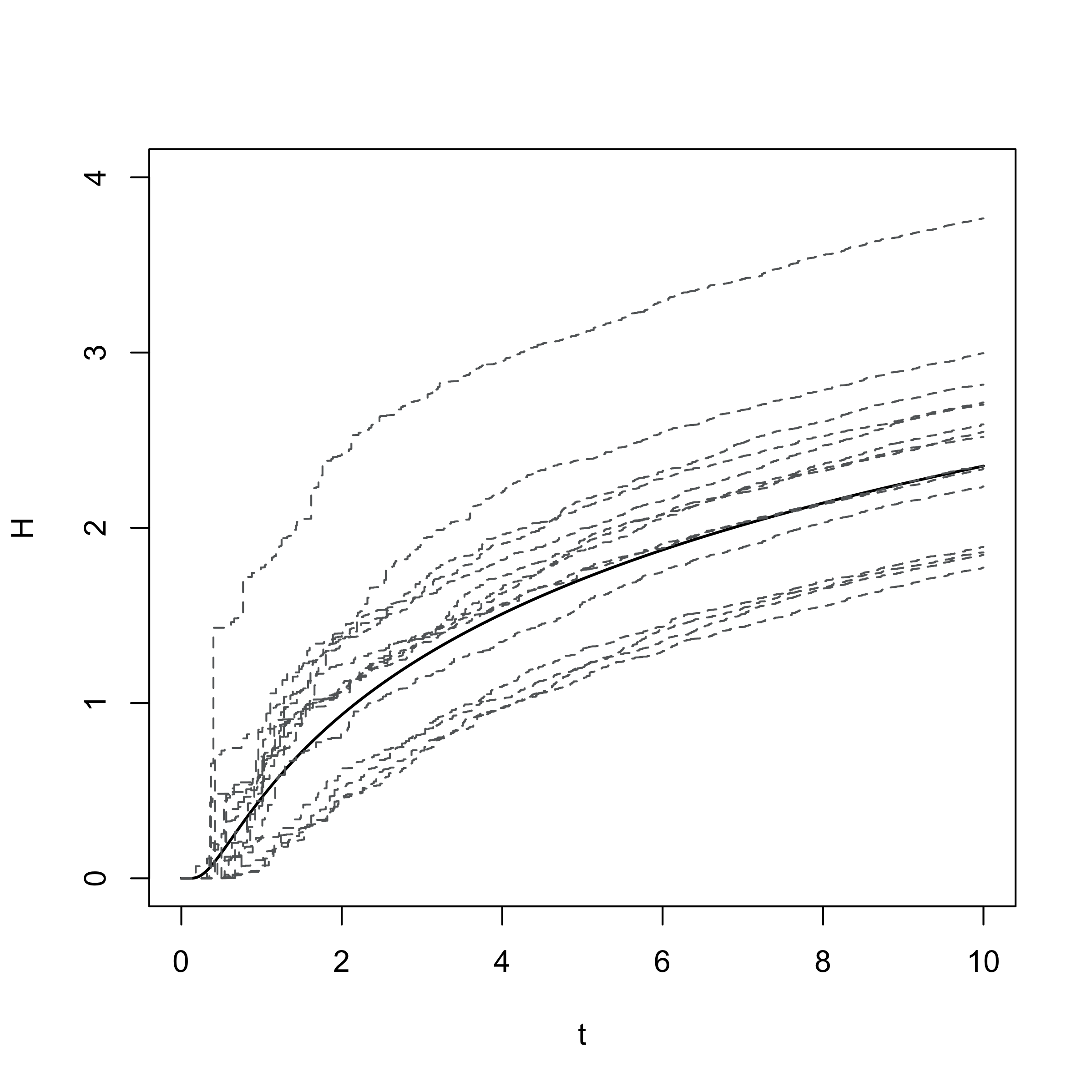}
    \caption{Simulated prior (dotted) and true (solid) curves for the same cumulative hazard function parameter $\Lambda(t)=-\log(1-\exp(-t^{-1}))$ (corresponding to the Fr\'echet distribution), and different choices of $c(t)=100,\,10,\, 10\,t^{-2},\, 10\,t^{2}$, respectively.} 
    \label{fig:priors}
\end{figure}

\section{Asymptotic properties}

In this section we present the Bayesian consistency of our randomised models, as well as their asymptotic normality. We emphasise the following nature of these results: these limits hold on compact intervals as the size of the dataset increases. For finite samples and on unbounded intervals, the baseline (or prior) distribution dominates the behaviour of the posterior (with larger fluctuations occurring further away from $0$). This is precisely where the specification of our baseline distribution via a different estimation mechanism comes into play in Section~\ref{sec:5} below.

\subsection{Bayesian consistency}\label{sec:3a}
Consistency for $H$ or $F$ is equivalent, since they are linked through continuous functionals. In particular, if we denote by $\Pi_n(\cdot)=\pr(\cdot|\nu_n^H,\mathcal{D}_n)$ the posterior distribution, we say that $H$ is \emph{consistent} in the posterior distribution $\Pi_n$  if for every $\varepsilon>0$, as $n\to\infty$,
\begin{align*}
\Pi_n\bigg(H: \,\sup_{t\in [0,\tau]}|H(t)-H_0(t)|>\varepsilon\bigg)\to0, \quad \tilde{\mathcal{L}}_\infty -\mbox{a.s.},
\end{align*}
for some $\tau$ in the interior of the support of the survival and censoring supports, where $\tilde{\mathcal{L}}_\infty$ is the joint law of all  $\nu_n^H$ and $\mathcal{D}_n$, and where $H_0$ is the \emph{true} (deterministic) cumulative hazard function. Likewise, we denote by $F_0$ and $G_0$ the true distribution functions of $T_i$ and $C_i$, respectively. Bayesian consistency for $F$ is very similar to the classical case.

\begin{theorem}[Consistency of $H$]\label{thm:consist}
Let the prior on $H$ be a conditional L\'evy process with intensity random measure of the form
\begin{align*}
\nu_n^H(\dd t,\dd x)=q_n(t,x)\Lambda_n(\dd t)\frac{\dd x}{x}.
\end{align*}
Assume that $\mathcal{L}_n-\mbox{a.s.}$ the following hold: $\Lambda_n$ are continuous cumulative hazards, $q_n$ are weakly continuous in the first component, and
\begin{equation*}
Q_n(\tau)\coloneqq\sup_{x\in(0,1)}\sup_{t\in[0,\tau]} (1-x)^{\kappa_n}q_n(t,x)<\infty,\quad
\sup_{u\in(0,\ve_n]}\sup_{t\in[0,\tau]}\bigg|\frac{q_n(t,u)}{q_n^0(t)}-1\bigg|\stackrel{n\to\infty}{\to}0,
\end{equation*}
$\tilde{\mathcal{L}}_\infty$-a.s., for some positive random processes $q_n^0$ and $Q_n$ and positive finite random variables $\kappa_n$ and $\ve_n\in(0,1)$ satisfying\footnote{We say that $a_n=o(b_n)$ (resp. $a_n=\omega(b_n)$) if $\lim_{n\to\infty}a_n/b_n$ equals $0$ (resp. $\infty$).} $\ve_n=o(1)$ and $\ve_n=\omega(n^{-1}\log n)$ $\tilde{\mathcal{L}}_\infty$-a.s. Then, the posterior distribution of $H$ is consistent on $[0,\tau]$ whenever
\begin{align}\label{convergence_kappa}
n^{-1}[\kappa_n+Q_n(\tau)(\Lambda_n(\tau)\vee 1)]\stackrel{\pr-a.s.}{\to}0,
\quad\text{as }n\to\infty.
\end{align}
\end{theorem}
We immediately obtain, from the continuity of the product integral transformation, that consistency of $H$ implies that of $F$.
\begin{corollary}[Consistency of $F$]
Let $F$ be constructed with $H$ satisfying the conditions of Theorem~\ref{thm:consist}. Then $F$ is consistent.
\end{corollary}

\begin{remark}
Some observations regarding the above result.
\begin{enumerate}
\item Equation \eqref{convergence_kappa} indirectly stipulates regularity conditions on the $q_n$ functions, which are easily satisfied. For instance, for a Beta process, the variables can be taken as $\kappa_n=1-\underline c_n\le 1$, where $\underline c_n=\inf_{t\in [0,\tau]} c_n(t)$ and where, for every $\tau\in(0,\infty)$ in the interior of the survival and censoring supports, $Q_n(\tau)=\sup_{t\in[0,\tau]}c_n(t)$ should satisfy $Q_n(\tau)(\Lambda_n(\tau)\vee 1)=o(n)$ a.s. It is easy to see that $q_n^0=c_n$ and that
\begin{equation}
\label{eq:growth_bound}
\begin{split}
\sup_{u\in(0,x]}\sup_{t\in[0,\tau]}\Big|\frac{q_n(t,u)}{q_n^0(t)}-1\Big|
&=\sup_{t\in[0,\tau]}\big|(1-x)^{c_n(t)-1}-1\big|\\
&\le\max\big\{(1-x)^{\underline{c}_n-1},Q_n(\tau)\big\}x.
\end{split}
\end{equation}
Thus, $c_n$ may be chosen to be convergent to $0$ (arbitrarily fast) in some regions or divergent in others (with a sufficiently slow divergence). Indeed, if $\Lambda_n(\tau)=O_\pr(1)$ as $n\to\infty$, it suffices to have the following limits $\tilde{\mathcal{L}}_\infty$-a.s. as $n\to\infty$: $Q_n(\tau)=o(n)$ and $%\sup_{t\in[0,\tau]}\big|(1-\ve_n)^{c_n(t)-1}-1\big|\le
\max\big\{(1-\ve_n)^{-1},Q_n(\tau)\big\}\ve_n\to 0$, for \emph{some} $\varepsilon_n=\omega(n^{-1}\log n)$. Thus, it suffices to assume that $Q_n(\tau)=o(n/\log n)$ $\tilde{\mathcal{L}}_\infty$-a.s.
\item The fact that we should choose $\tau$ in the interior of the support of the survival and censoring supports is crucial to the proof. This restriction can be relaxed in the frequentist case, see \cite{Wang1987}.
\end{enumerate}
\end{remark}

\subsection{Bernstein--von Mises theorems}\label{sec:3b}

In what follows, convergence in distribution of processes is understood in the sense of random elements in the Skorokhod space $\mathbb{D}([0,\tau])$ of functions which are right continuous with left limits, equipped with the uniform norm. Let $B$ standard Brownian motion and $U_0=\int_{[0,\cdot)}(\bar{F}_{0}\bar{G}_{0})_-^{-1}\dd H_0$.
\begin{theorem}[Bernstein--von Mises for $H$]\label{thm:BvM1}
Let the prior on $H$ be a conditional L\'evy process with intensity random measure of the form
\begin{align*}
\nu_n^H(\dd s,\dd x)=q_n(s,x)\dd s\frac{\dd x}{x}.
\end{align*}
Assume that $\mathcal{L}_n-\mbox{a.s.}$ the following hold: $q_n$ are continuous in the first entry, and
\begin{align*}
Q_n(\tau)\coloneqq\sup_{s\in[0,\tau],\,x\in(0,1)} (1-x)^{\kappa_n}q_n(s,x)<\infty,
\end{align*}
for some random processes $q_n^0$ and positive random variables $\kappa_n$. Then:
\begin{enumerate}
\item If $\kappa_n +Q_n(\tau)(\Lambda_n(\tau)\vee 1)=o_\pr(n)$ $\tilde{\mathcal{L}}_\infty$-a.s., then
\begin{align*}
\sqrt{n}\big(H|(\nu_n^H,\mathcal{D}_n) -\E_{\nu_n^H,\mathcal{D}_n} H\big)\stackrel{d}{\to} B\circ U_0,\quad \tilde{\mathcal{L}}_\infty-\mbox{a.s.}
\end{align*}
\item If for some $\eta\in\mathbb{R}$, there exist random variables $C_n$, $\ve_n=o(1)$ (on $(0,1)$) with $\ve_n=\omega(n^{-1}\log n)$ and $\alpha\in(1/2,1] $, such that
\begin{align*}
\sup_{x\in[0,\ve_n]}\sup_{t\in[0,\tau]}|q_n(t,x)/q_n^0(t)-1|\le C_n \ve_n^{\alpha},
\end{align*}
then we have
\begin{align*}
\sqrt{n}\big(\E_{\nu_n^H,\mathcal{D}_n} H-H_n\big)=O_\pr([1\vee C_n]n^{-\alpha+1/2}),\quad \tilde{\mathcal{L}}_\infty-\mbox{a.s.}
\end{align*}
In particular the term vanishes if $C_n=o_\pr(n^{\alpha-1/2})$.
\item
Consequently, under both assumptions, for $C_n=o_\pr(n^{\alpha-1/2})$,
\begin{align*}
\sqrt{n}\big(H|(\nu_n^H,\mathcal{D}_n)-H_n\big)&\stackrel{d}{\to} B\circ U_0,\quad \tilde{\mathcal{L}}_\infty-\mbox{a.s.}\\
\sqrt{n}\big(\E_{\nu_n^H,\mathcal{D}_n} H-H_0\big)&\stackrel{d}{\to} B\circ U_0.
\end{align*}
\end{enumerate}
\end{theorem}

By applying the functional $\delta$-method using the product integral, which is a Hadamard differentiable map, we may obtain an analogous result for the survival function.
\begin{corollary}[Bernstein von--Mises for $F$]\label{cor:cvmF} Let $F$ be constructed with $H$ satisfying all the conditions of Theorem~\ref{thm:BvM1}. Then 
\begin{align*}
\sqrt{n}\big(F|(\nu_n^H,\mathcal{D}_n)-F_n\big)&\stackrel{d}{\to} F_0\, B\circ U_0,\quad \tilde{\mathcal{L}}_\infty-\mbox{a.s.}\\
\sqrt{n}\big(\E_{\nu_n^H,\mathcal{D}_n} F-F_0\big)&\stackrel{d}{\to} F_0\, B\circ U_0.
\end{align*}
\end{corollary}

\begin{remark}
    We make the following observations related to the above theorem.
    \begin{itemize}
\item We essentially require the L\'evy measure of $A=-\log\ov F$ to be temporally inhomogeneous and its behaviour at $0$ is uniformly similar to that of a (time-inhomogeneous) gamma process. The assumptions on the characteristics $q_n$, $q_n^0$, $Q_n$ and $\Lambda_n$ all serve to quantify this dissimilarity. 
\item  Different prior specifications yield different posterior means, which can be compared against frequentist models, such as the Nelson--Aalen estimator of the cumulative hazard function or the Kaplan--Meier estimator of the survival function. The second result of Theorem~\ref{thm:BvM1} gives conditions under which these posterior means behave asymptotically equivalently to frequentist models. Then, for instance, the second equation of Corollary~\ref{cor:cvmF} can be considered as a Donsker theorem for the posterior mean. Such an approach is the basis for non-parametric spliced models introduced below. 
    \end{itemize}
\end{remark}

\section{Exact simulation}\label{sec:4}

Most simulation algorithms of additive processes (such as the ones presented in~\citealp[Ch.~13]{ghosal2017fundamentals}) are approximate algorithms based on the representation of a beta process through a counting measure. More precisely, $H(t) = \sum_{s\in(0,t]}\Delta H(s)$, where for $t \le \tau$ the sum has countably many terms given by the points $(s,\Delta H(s))$ of a Poisson process on $(0, \tau] \times (0, 1)$ with mean measure given by $\nu_H^d+\nu_H^c$, where $\nu_H^d$ is the discrete component whose jumps times are fixed and $\nu_H^c$ is the continuous part, and both admit the representation
\begin{align*}
\nu_H^c(\dd t, \dd x) 
&= b(t)(1-x)^{b(t)-1} \frac{\dd x}{x} \dd\Lambda^c(t),\\
\nu_H^d 
&= \sum_{t:\Delta\Lambda^d(t)\ne 0}\delta_{(t,\Be(b(t)\Delta\Lambda^d(t),b(t)(1-\Delta\Lambda^d(t))))},
\end{align*}
where $b:[0,\infty)\to [0,\infty)$ is measurable, $\Lambda^c$ is continuous and non-decreasing and $\Lambda^d$ is non-decreasing and piecewise constant with $J\in\mathbb{N}$ jumps. In other words, the process $H$ admits the decomposition $H=H^c+H^d$, where $H^c$ is an additive process with L\'evy measure $\nu_H^c$ independent of $$H^d(t)=\int_{(0,t]\times\R_+}x\,M^d(\dd s,\dd x),$$ where $M^d=\sum_{i=1}^{J}\delta_{(t_i,\xi_i)}$ and $t_1,\ldots,t_J$ are the jump times of $\Lambda^d$ and the variables $\xi_i$ are mutually independent with $\xi_i\sim\Be(b(t_i)\Delta\Lambda^d(t_i),b(t_i)(1-\Delta\Lambda^d(t_i)))$. 

The main drawback of all approximate simulation methods in the literature is the lack of control over the error of the approximations for both, the weak error (e.g., bias) and the strong error (i.e., pathwise). This control is absent for the hazard $H$, the survival probability $\ov{F}=1-F$, and its logarithm $A=-\log \ov{F}$. Since such approaches would require infinite computational power to sample the entire path of either $H$ or $\ov{F}$, in this paper, we consider the \emph{exact} simulation of the \emph{marginals} of these processes along a specified arbitrary sequence of times. The advantages of exact simulation algorithms over approximate methods are well documented and particularly valuable when one is interested in boundary (or limit) behaviour as well as extreme events (see e.g.~\citealp[\S3.3.1]{GCFirstPassage}). 

In the remainder of this section, we assume that $H$ denotes the posterior process corresponding to a continuous baseline hazard. In other words, by virtue of~\citep[Thm~13.5]{ghosal2017fundamentals} we assume that, for some continuous $\Lambda_0$, we have  
\[
b(t)=c(t)+Y(t),\quad
\dd\Lambda^c(t)=\frac{c(t)}{b(t)}\dd\Lambda_0(t)
\quad\text{and}\quad
\dd\Lambda^d(t)=\frac{1}{b(t)}\dd N(t).
\]

\subsection{Simulation of the survival probability}
\label{subsec:sample_A}

The process $A=-\log \ov{F}$ is an additive process that can be written as the sum of independent additive processes $A^c$ and $A^d$, with respective L\'evy measures (see~\citealp[Prop.~13.10]{ghosal2017fundamentals}) $\nu_A^d$ and $\nu_A^c$ and satisfying 
\[
\nu_A^c(\dd t, \dd x) 
= c(t)e^{-b(t)x} \frac{\dd x}{1-e^{-x}} \dd\Lambda_0(t),
\quad 
A^d(t)=-\sum_{i=1}^{N(\infty)}I(t_i\le t)\log(1-\xi_i),
\]
where $t_1<\ldots<t_{N(\infty)}$ are the jump times of $N$ and $\xi_i\sim\Be(\Delta N(t_i),b(t_i)-\Delta N(t_i))$ are independent for $i\in\{1,\ldots,N(\infty)\}$. Moreover, we assume that $c$ is piecewise constant, implying that so is $b$. Thus, for some $0=\beta_0<\ldots<\beta_{K}<\beta_{K+1}=\infty$, we have $b(t)=\sum_{i=1}^{K+1} b(\beta_i)I(\beta_{i-1}<t\le \beta_i)$ (and $c$ is constant on the intervals $(\beta_i,\beta_{i+1}]$).

Since the jumps of the discrete L\'evy measure $\nu^d_A$ are easy to simulate (they are fixed in time by the data and the jump sizes are easily sampled), we concentrate on sampling the additive process $A^c$. We next present an algorithm that simulates the marginal $A^c(\tau)$ for any fixed $\tau>0$. Since $A^c$ has independent increments, sampling such marginals is sufficient to produce samples of its finite dimensional laws.

%Define the global bound $\ov{b}=\max_{i\in\{0,\ldots,K\}}b_i$ on $b$. 
The idea is to write $A^c$ as the sum of two independent processes $A^c=B+C$ where $B$ is a time-changed gamma process and $C$ is a time-inhomogeneous compound Poisson process. Since $B$ has exact simulation methods and $C$ has finitely many jumps that may be simulated exactly with relative ease, we obtain an exact simulation scheme for $A$. To this end, let $B$ and $C$ be independent additive processes with L\'evy measures given by 
\begin{gather*}
\nu_B(\dd x,\dd t)
=e^{-b(t)x}\frac{\dd x}{x}c(t)\dd\Lambda_0(t),
\quad
\nu_C(\dd x,\dd t)
=e^{-b(t)x}
\phi(x)%\frac{e^{-x}-1+x}{x(1-e^{-x})}
\dd x\,c(t)\dd\Lambda_0(t),%\\
%\text{where}
%\enskip
%\phi(x)
%=\frac{e^{-x}-1+x}{x(1-e^{-x})}.
%\quad\text{and}\quad
%\phi_2(a,x)
%=\frac{1-e^{-ax}}{1-e^{-x}}.
\end{gather*}
where $\phi(x)
=(e^{-x}-1+x)/(x(1-e^{-x}))$. It is easy to verify that, if $G_i$ are iid standard gamma processes (i.e., L\'evy processes with L\'evy measure $e^{-x}x^{-1}\dd x$ and marginals $G_i(t)$ having the density $x\mapsto e^{-x}x^{t-1}/\Gamma(t)$), then 
\begin{align*}
(B(t);t\ge 0)
&\eqd 
\bigg(\int_0^t \frac{1}{b(t)}\dd G_1\bigg(\int_0^t c(s)\dd\Lambda_0(s)\bigg);t\ge 0\bigg)\\
&=\bigg(\sum_{i=1}^{K+1}\frac{G_i\big(c(\beta_i)(\Lambda_0(\beta_i\wedge t)-\Lambda_0(\beta_{i-1}\wedge t))\big)}{b(\beta_i)};t\ge 0\bigg),
\end{align*}
which is easily simulated. 

On the other hand, $C$ has finite activity as $\nu_C(\R_+,(0,\tau])<\infty$. The process $C$ can be simulated easily via thinning. Indeed, consider an additive process $C^*(t)$ with L\'evy measure $\nu_{C^*}(\dd x,\dd t)=e^{-b(t)x}\dd x\, c(t)\dd\Lambda_0(t)$. Thus, we may sample $C^*$ as in~\cite[Ch.~13]{ghosal2017fundamentals}, but in finitely many steps and then thin the jumps of $C^*$ to obtain the jumps of $C$. Indeed, we may sample the epochs of $C^*$ as the epochs of a standard Poisson process time-changed with the function 
\[
t\mapsto \nu_{C^*}(\R_+,(0,t])
= \int_0^t\frac{c(s)}{b(s)}\dd\Lambda_0(s)
= \sum_{i=1}^{K+1}\frac{c(\beta_i)}{b(\beta_i)}(\Lambda_0(\beta_i\wedge t)-\Lambda_0(\beta_{i-1}\wedge t)). 
\]
Conditionally given the epochs of $C^*$ on $(0,\tau]$, say $T_1<\cdots<T_{M(\tau)}$, the corresponding jump sizes $J_1,\ldots,J_{M(\tau)}$ of $C^*$ are independent exponential variables with corresponding rates $b(T_1),\ldots,b(T_{M(\tau)})$, which can be easily sampled. Then, we may accept each jump pair $(T_n,J_n)$ as a jump of $C$ independently with probability $\phi(T_n)\in(1/2,1)$. Finally, we note that $C(\tau)$ is equal to the sum of all accepted jump sizes.

Given the exact simulation of $A=-\log \ov{F}$, we may easily sample the marginals of $\ov{F}$ as well, since $\ov{F}=\exp(-A)$. This can be used, in turn, to obtain Monte Carlo estimators for, say, the median tail integrals $\int_t^\infty\ov{F}(s)\dd s$ at arbitrary times $t$. We may also construct credible regions: asymptotic intervals can be produced via the Bernstein--von Mises result in Theorem~\ref{thm:BvM1} while non-asymptotic credible regions may be constructed via concentration inequalities such as Hoeffding's inequality or Chernoff's bound. This methodology also leads us closer to the exact simulation of the random quantiles of $\ov{F}$, that is, the random first passage time of $\ov{F}$ across some level $q\in(0,1)$ or, equivalently, the first time $A$ hits some level $-\log q>0$. Indeed, as described in Remark~\ref{rem:hitting_time} below, this would be possible if we were able to generate exact samples for the hitting time of a standard gamma process, which unfortunately does not exist yet.

\begin{remark}
\label{rem:hitting_time}    
Indeed, to simulate the first hitting time of $A$ across a level $-\log q$ where $q\in(0,1)$ is a quantile, we can do the following: discretise time into intervals $(0,\tau_1],\,(\tau_1,\tau_2],\,\ldots$ where we include the jump-times of $C$ and those of $A^d$ until $C+A^d$ cross the desired level (the total number of jumps is stochastically dominated by a geometric random variable). We would then sample the increments of $B$ over these intervals until we identify the interval $(\tau_{n-1},\tau_n]$ during which $A$ crossed level $-\log q$. If the crossing happened at some time $\tau_n$ exactly (i.e., because of a jump of $C$), the procedure is over. If instead it occurred during an interval $(\tau_{n-1},\tau_n)$, then we drop the increment of $B$ and simply sample the time and value of process $B$ conditioned to hit the `right' level on that interval: $B_t-B_{\tau_{n-1}}\ge -\log q - A_{\tau_{n-1}}$ for some $t\in(\tau_{n-1},\tau_n)$. Since $B$ is a time-changed gamma process, it would suffice to have an algorithm that simulates the first passage time of a gamma process across a fixed level. Unfortunately, this algorithm does not yet exist and the available first passage simulation algorithms cannot be directly adapted to this process (see, e.g.~\citealp{MR4122822,GCFirstPassage2,GCFirstPassage}).
\end{remark}

\subsection{Simulation of the hazard function}
\label{subsec:sample_H}

To draw exact samples of the marginals of $H$, we may follow a similar methodology, as we now describe. Our goal is to decompose $H$ into the sum of independent processes, one of which we sample elementarily, another that can be simulated by time-changing a tempered Dickman process (also known as a truncated gamma process) and another process that results by thinning a compound Poisson process. To be precise, consider the decomposition $H=H^d+D+E$, where $H^d$ is as before and can be sampled easily via beta random variables, $D$ is a piecewise tempered Dickman process and $E$ is a compound Poisson process with respective L\'evy measures
\begin{align*}
\nu_D(\dd t,\dd x)
&=c(t)\frac{e^{-2\log(2)(b(t)-1)^+x}}{x}I(x\le 1/2)\dd x \dd\Lambda_0(t),\\
\nu_E(\dd t,\dd x)
&=c(t)\frac{(1-x)^{b(t)-1}-e^{-2\log(2)(b(t)-1)^+x}I(x\le 1/2)}{x}I(x\le 1)\dd x \dd\Lambda_0(t),
\end{align*}
and where $H^d$, $D$ and $E$ are independent. To sample the process $E$ we may sample the jump times and sizes of a compound Poisson process $E^*$ with L\'evy measure
\[
\nu_{E^*}(\dd t,\dd x)
=2c(t)\big[\psi(x,b(t)) \,I(x\le 1/2)
    +(1-x)^{b(t)-1}\,I(1/2<x<1)\big]\dd x \dd\Lambda_0(t),
\]
where 
\[
\psi(x,y)=(2^{1-y}-1)\,I(y\le 1) + (\log 2-1/2)(y-1)(1-x)^y\,I(y>1),
\]
and then thin the jumps of $E^*$ to obtain the jumps of a process with the law of $E$. We note here that $E^*$ is again a compound Poisson process, whose jumps smaller than $1/2$ are uniformly distributed when $b(t)\le 1$ and otherwise have the same law as the random variable $1-(1-(1-2^{-b(t)-1})U)^{1/(b(t)+1)}$ for $U\sim \mathrm{U}(0,1)$, while the jumps larger than $1/2$ have the same law as $1-U^{1/b(t)}/2$. 

The process $D$ can be sampled similarly to $B$, via~\cite[Algs.~3.1~\&~3.2]{MR3981146}. Indeed,~\cite[Alg.~3.4]{MR3981146} samples from a subordinator $L$ at time $t$ with L\'evy measure given by $I(x\le 1)e^{-\mu x}x^{-1}\dd x$, implying that the L\'evy measure of $L_t/2$ is given by $I(x\le 1/2)e^{-2\mu x}x^{-1}\dd x$. Thus, on every interval of constancy $[t,u]$, setting $\mu=\log(2)(b(t)-1)^+$, we have 
\[
\frac{L_{\Lambda(t+\Delta t)}-L_{\Lambda(t)}}{2}
\eqd D_{t+\Delta t}-D_t,
\quad \Delta t\in[0,u-t].
\]
In principle, the cited algorithm requires a numerical optimisation of certain hyper-parameters prior to simulations, which depends on $\mu$ (and hence on $b(t)$). Thus, this would need to be done every time $b$ takes a different value, which is impractical. We propose picking the hyper-parameters following the simple rule $\vartheta=(1+\mu)^{-1}$ and $\delta=1-\log(e^2+\mu)^{-1}$ in~\cite[Alg.~3.2]{MR3981146}, which appears to be a choice that keeps the acceptance probability reasonably large as a function of $\mu$ (see Subsection~\ref{subsec:Dassios-constant} below for details). 

\subsection{Simulation of the truncated gamma process}
\label{subsec:Dassios-constant}

Consider a truncated gamma process with rate $\mu$. An elementary analysis into~\cite[Alg.~3.2]{MR3981146} reveals that the expected running time of this algorithm is proportional to $(t+1)/C(\vartheta,\delta,\mu)$, where $t$ is the time at which this process is sampled and $1/C(\vartheta,\delta,\mu)$ is the acceptance probability of an acceptance-rejection step in the algorithm, given by the formula
\begin{align*}
&C(\vartheta,\delta,\mu) 
= e^{-\mu-1}\frac{\Gamma(\delta)}{1-\delta}\cdot\frac{\exp(\zeta e^\zeta)}{\vartheta\Gamma(e^\zeta+\delta)},\\
&\enskip\text{where}\enskip
\zeta = \Gamma(0,\mu)+\vartheta+\log\mu,
\enskip \vartheta>0,
\enskip\delta\in(0,1),
\end{align*}
and $\Gamma(0,\mu)=\int_\mu^\infty t^{-1}e^{-t}\dd t$ is the upper incomplete gamma function. Given the complex dependence on the free parameters $\vartheta>0$ and $\delta\in(0,1)$, the authors of~\cite{MR3981146} suggest numerically optimising this constant. However, since our parameters are variable as a function of time, as described above, we would have to do this optimisation repeatedly. To circumvent this issue, we opt for a simple and reasonable rule of thumb that can be easily computed. The analysis found in the Supplement yields the following simple choice: $\delta=1-1/\log(e^2+\mu)$ and $\vartheta=(1+\mu)^{-1}$.

%\subsection{Posterior approximation error bounds}
\section{Constructing non-parametric spliced models}\label{sec:5}

When modeling real-world data, a significant focus has emerged regarding the concept of parametric splicing, which effectively means having different parametric families in the body and tail of the distribution. Heavy-tailed losses in insurance are one main example where such an accurate tail description is imperative, see \cite{reynkens2017modelling}. This section provides Bayesian-based estimators derived as the posterior mean of survival curves, where the tail behaviour of the data is captured through a stochastic hyperparameter. The stochasticity arises from the fact that a Hill-type estimator for the tail is used. The procedure can thus be regarded to some extent as belonging to Empirical Bayes techniques.

\subsection{Regularly-varying tails}
Assume that the distribution is regularly-varying in the tail (cf. \citealp{bingham1989regular}), which is a broad heavy-tailed class of distributions with survival functions of the form $\ell(t) t^{-\alpha}$, where the function $\ell$ is slowly-varying at infinity, which by definition means that $\lim_{s\to\infty}\ell(as)/\ell(s)=1$ for all $a>0$. Loosely speaking, these distributions are Pareto in the tail. Whether censored data falls into this tail regime can be checked using the so-called Pareto QQ-plot, given by
$$\big(\log t , \,  -\log (1- \hat{F}_0(t)) \big)$$
where $\hat{F}_0$ is a suitable first-step estimator of the distribution $F_0$, for instance, the (frequentist) Kaplan--Meier estimator. The plot should be linear for large $t$. Alternatively, prior knowledge can directly indicate that the data falls into this tail regime.

Next consider a Beta L\'evy distribution for the prior distribution with parameters $(c_n,\Lambda_n)$. A natural choice for the tail behaviour of the prior is given by
\begin{align*}
\frac{\dd}{\dd t}\Lambda_n(t)=\frac{\alpha}{t},
\quad\text{for large }t,
\end{align*}
where $\alpha$ is to be replaced by a suitable estimator $\hat\alpha_{k,n}$ of the tail parameter under censoring. Let the order statistics of the sample be denoted by
$$T_{1,n}\le T_{2,n}\le\cdots\le T_{n,n},$$
and $\delta_{[j,n]}$ the concomitant of the order statistic $T_{j,n}$.
Then these may be obtained from \cite{beirlant2007estimation} or \cite{bladt2024censored} (see also a third alternative in~\citealp{bladt2021trimmed}), given respectively by
\begin{gather}
\label{eq:Hill}
\hat\alpha_{k,n}=\frac{\sum_{j=1}^k \delta_{[n-j+1,n]}}{\sum_{j=1}^k \log(T_{n-j+1,n}/T_{n-k,n})},\\
\nonumber
\hat\alpha_{k,n}=\frac{\sum_{j=1}^k \omega_{jk}}{\sum_{j=1}^k \omega_{jk}\; \log(T_{n-j+1,n}/T_{n-k,n})},\: \omega_{jk} = \frac{\delta_{[n-j+1:n]}}{j} \prod_{l=j+1}^k \left[\frac{l-1}{l}\right]^{\delta_{[n-l+1,n]}}.
\end{gather}
The estimators may also be defined using any other data, or expert information. The heavy-tailed rate should only be prescribed on from a certain point where the power-law decay of the survival function is reasonable. Hence, as a mechanism of shutting the prior on-and-off, we may for instance specify the second parameter as
\begin{equation}
\label{eq:c_n-a_n}
c_n(s)
=2^{-n}I(s<T_{n-k,n})+a_n I(s\ge T_{n-k,n}),
\end{equation}
where $a_n$ is an increasing sequence modulating the precise proportions that the posterior distribution assigns to the data and the prior, respectively. 

The precise speed at which $a_n$ is allowed to grow is given in Theorem~\ref{thm:BvM1}, and all asymptotic results follow. Indeed, by~\eqref{eq:growth_bound}, we may pick any $a_n=o(\sqrt{n})$. Of interest is the case of \textit{exact splicing}, which we define as the case where $a_n\equiv\infty$, which implies that, from the $k$-th largest observation onwards, the Kaplan--Meier behaviour is shut off completely and fully replaced by a Hill-type estimated Pareto prior. 

From \eqref{posterior_beta} we have that for a conditional Beta process, the posterior mean and variance are 
\begin{align*}
\E_{\tilde\nu_n^H} H(t)&=\int_{(0,t]}\frac{c_n\dd \Lambda_n+\dd N_n}{c_n+Y_n}\\
\mathrm{var}_{\tilde\nu_n^H} H(t)&=\int_{(0,t]}\left(1-\Delta\frac{ N_n}{c_n+Y_n}\right)\frac{c_n\dd \Lambda_n+\dd N_n}{(c_n+Y_n)(c_n+Y_n+1)}.
\end{align*}

Expanding the terms yields novel \emph{non-parametric spliced} estimators for the hazard and survival functions in heavy-tailed settings:
\begin{align*}
\Lambda_{k,n}(t)&=\hat\alpha_{k,n} \int_1^t \frac{c_n(s)}{c_n(s)+Y_n(s)} \frac{\dd s}{s} + \int_0^t \frac{1}{c_n(s)+Y_n(s)} \dd N_n(s)\\
\overline F_{k,n}(t)&=\prodi_0^t (1-\dd \Lambda_{k,n}(s))\\
&=\exp\bigg(- \hat\alpha_{k,n} \int_1^t \frac{c_n(s)}{c_n(s)+Y_n(s)}\frac{\dd s}{s}\bigg)\prod_{s\in(0,t]}\bigg(1-\frac{\Delta N_n(s)}{c_n(s)+Y_n(s)}\bigg).
\end{align*}
Observe that $Y_n$ is a non-decreasing process that vanishes above the largest observation, and the exponential term becomes a pure Pareto tail. The product term guarantees that the Pareto tail is ``pasted" at the correct location, that is where the product-limit estimator left off. This pasting is made in a gradual manner thanks to the sequence $c_n$ which regulates the intensity of each of the components below the largest observation.

We now specialise to the case with no censoring. In that case we have that both estimators $\hat\alpha_{k,n}$ reduce to the classic Hill estimator
\begin{align*}
\hat\alpha_{k,n}
&=k\left[\sum_{j=1}^k \log(T_{n-j+1,n}/T_{n-k,n})\right]^{-1},
\qquad\text{leading to}\\
\Lambda_{k,n}(t)&=\hat\alpha_{k,n} \int_1^t \frac{c_n(s)}{c_n(s)+Y_n(s)} \frac{1}{s} \dd s + \sum_{i=1}^n \frac{I(T_i\le t)}{c_n(T_i)+\sum_j I(T_j\ge T_i)}
\\
\overline F_{k,n}(t)
&=\exp\bigg(- \hat\alpha_{k,n} \int_1^t \frac{c_n(s)}{c_n(s)+Y_n(s)}\frac{\dd s}{s}\bigg)\prod_{j:\,T_j\le t}\bigg(1-\frac{1}{c_n(T_j)+Y_n(T_j)}\bigg).
\end{align*}
The above formula shows an intricate combination of the non-parametric empirical distribution function (EDCF) and a pure Pareto tail. It is straightforward to show that as long as $a_n$ is a diverging sequence, the asymptotic behaviour of these spliced models follow those of the Kaplan--Meier estimator and ECDF, respectively.

\subsection{Weibull tail behaviour}
\label{subsec:weibull}

In some applications, the tail behaviour can be correctly captured through Weibull-type tails, given by $e^{-t^{p} \ell (t)}$ where $p>0$ is a parameters controlling the shape of the distribution, and where $\ell$ is a slowly-varying function as described above. This tail regime can be verified through the linearity for large $t$ of the Weibull-QQ plot, given by
$$
\big( \log(t), \log\{-\log (1-{\hat{F}_0(t)})\}\big),
$$
where again $\hat{F}_0$ is a suitable first-step estimator of the distribution $F_0$, such as the Kaplan--Meier estimator or, if the true distribution is assumed to be continuous, $-\log(1-\hat{F}_0)$ can be replaced with the Nelson--Aalen estimator. Prior knowledge can also directly classify into this tail regime.

Taking again a Beta L\'evy distribution with parameters $(c_n,\Lambda_n)$, we may specify
\begin{align*}
\frac{\dd}{\dd t}\Lambda_n(t)
=\alpha t^{p-1}
=\frac{p}{l^p}\,t^{ p-1}
\quad\text{for large }t,
\end{align*}
where $p$ is a parameter that regulates the tail regime and $l$ is a scale parameter. For instance $p=1$ implies an exponential decay of the tail, while $0<p<1$ is the sub-exponential case (sometimes called \emph{stretched exponential}), with $p\downarrow0$ nearing (but not reaching) Pareto-type tails. Similarly, the case $p>1$ is the super-exponential case, implying very light tails. In practice, these parameters should be replaced by suitable estimators $\hat p_{k,n}$ and $\hat l_{k,n}$.

The estimators of the shape and scale parameters can be deduced from the Weibull QQ-plot, since the theoretical slope of the plot is precisely $p$. Thus we may set
\begin{align*}
&SS_w(p,l)=\\
&\sum_{j=1}^k\left( 
\log \big(%\frac{
-\log(1-\hat F_0(T_{n-j+1,n}))%}{\log(1-\hat F_0(T_{n-k,n}))}
\big)
-p \log(T_{n-j+1,n}/T_{n-k,n})+p\log(l)
 \right)^2, 
\end{align*}
and then set
\begin{align*} 
(\hat p_{k,n},\hat l_{k,n})
\coloneqq\arg\min_{(p,l)} SS_w(p,l)
=\big(\mbox{cov}(x,y)/\mathrm{var}(x),\:\exp(\bar x-\bar y\,\mathrm{var}(x)/\mbox{cov}(x,y) )\big),
\end{align*}
with $x,y$ the vectors with entries
$$y_j=\log \big(%\frac{
-\log(1-\hat F_0(T_{n-j+1,n}))%}{\log(1-\hat F_0(T_{n-k,n}))}
\big),\quad x_j=\log(T_{n-j+1,n}/ T_{n-k,n}),\quad j=1,\dots, k.$$
Then with an appropriate choice of $c_n$ sequence, we obtain
\begin{align*}
&\overline F_{k,n}(t)=\\
&\exp\bigg(- \hat p_{k,n} \hat l_{k,n}^{-\hat p_{k,n}}\int_0^t \frac{c_n(s)}{c_n(s)+Y_n(s)} s^{\hat p_{k,n}-1}  \dd s\bigg)\prod_{s\in(0,t]}\bigg(1-\frac{\Delta N_n(s)}{c_n(s)+Y_n(s)}\bigg)
\end{align*}
for the censored case, which reduces in the uncensored case to
\begin{align*}
&\overline F_{k,n}(t)=\\
&\exp\bigg(- \hat p_{k,n} \hat l_{k,n}^{-\hat p_{k,n}} \int_0^t \frac{c_n(s)}{c_n(s)+Y_n(s)} s^{\hat p_{k,n}-1}  \dd s\bigg)\prod_{j:\,T_j\le t}\bigg(1-\frac{1}{c_n(T_j)+Y_n(T_j)}\bigg).
\end{align*}

It now becomes clear that other tail models beyond the Pareto-type and Weibull-type tails can be considered under the same framework, provided one can find a suitable non-parametric estimator for the parameter that regulates the tail. Also notice that we have provided estimators arising from the posterior mean, and that the full Bayesian posterior distribution can be used to provide credible regions for the above estimators, which is a significant advantage compared to the existing (parametric) spliced models in the literature.

\section{Numerical illustrations}\label{sec:6}

In this section, we present three examples of the non-parametric splicing at work, one inspired by the Pareto distribution, another inspired by the Weibull distribution, and a real-world survival data example. 

\subsection{Pareto-type synthetic data}\label{subsec:6-Pareto}
\begin{figure}[hbt!]
\centering{}
\includegraphics[width=6cm]{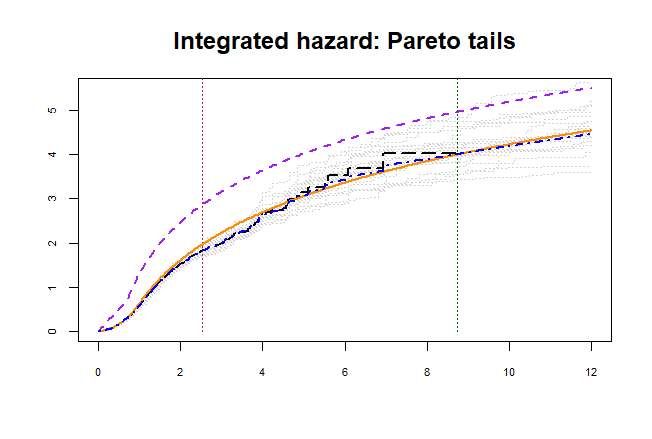}
\includegraphics[width=6cm]{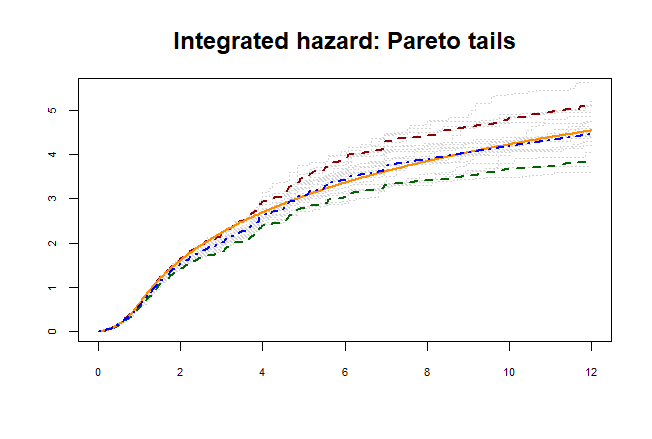}\\
\includegraphics[width=6cm]{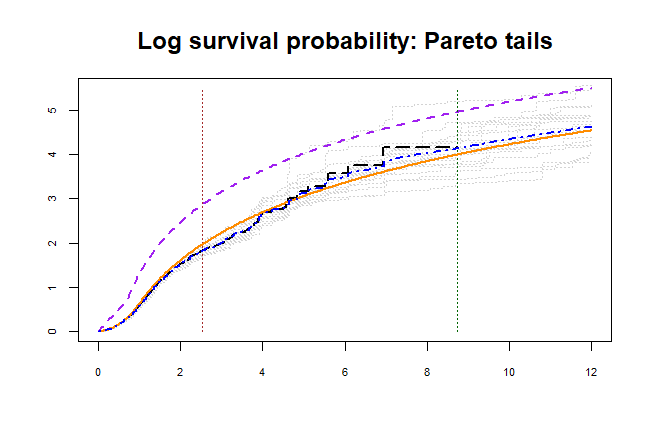}
\includegraphics[width=6cm]{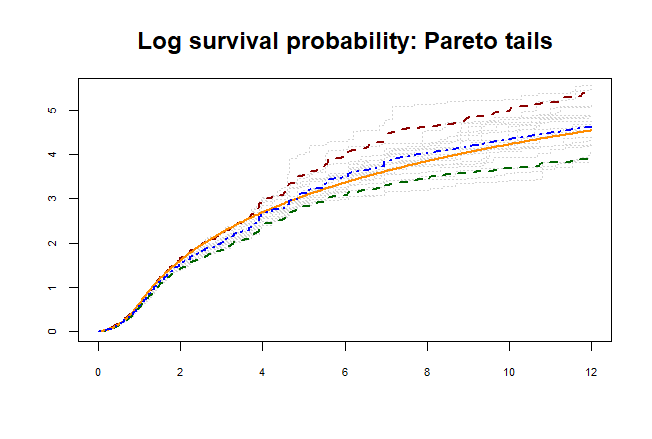}
\caption{Pareto tailed simulated data. We see 20 out of the 200 simulated paths dotted in light gray, the sample mean dot-dashed in blue, the prior dashed in purple, the Kaplan--Meier and Nelson--Aalen estimators long-dashed in black, the true hazard and log-survival solid in dark orange and, on the right, the upper and lower credible regions (generated via samples) long-dashed in red and green, respectively. The vertical dotted lines indicate where the $(n-k_n)$-th observation and where the largest observation are located.}
\label{fig:simulated-pareto}
\end{figure}

We consider simulated random variables, generated via the rule $X-U$, where $X$ is Pareto distributed with parameters $\alpha=1.8$ and $1$ (i.e. $\pr(X>t)=t^{-\alpha}$ for $t\ge 1$), $U$ is uniform on $(0,1)$, and both variables are independent. The censoring law is $1.4X-U$, where $X$ is Pareto distributed with parameters $(.7)\alpha$ and $1$ and $U$ is uniform on $(0,1)$, and both variables are independent. We draw $n=1000$ samples, assume from expert knowledge that our distribution is polynomially tailed (i.e., $p=0$) and use $a_n\coloneqq\log n$ and the quantile sequence $k_n\coloneqq \lceil2\sqrt{n}\rceil=64$ satisfying $k_n/n\to0$ and $k_n\to\infty$ as $n\to\infty$ for Hill's estimator of $\alpha$ as in~\eqref{eq:Hill} above, which yields the estimated value $\hat\alpha_n\approx 1.69$. This sharp estimation of the tail index informs our baseline hazard function, given by:
\[
\frac{\dd}{\dd t}\Lambda_n(t)
= qI\{t<t_0\} 
    + \frac{\hat \alpha_n}{t}I\{t\ge t_0\},
\quad t>0,
\]
for (arbitrarily chosen) $q=1$ and $t_0=T_{n-k,n}$. Further, we set $c_n$ via $a_n=\log n$ as in~\eqref{eq:c_n-a_n}. In Figure~\ref{fig:simulated-pareto}, we show a draw of 50 samples of the posterior and compare with the priors, the true hazard and log-survival functions $-\log\ov F(t)$, as well as the corresponding Kaplan--Meier and Nelson--Aalen estimators.

In the plots on the left-hand side of Figure~\ref{fig:simulated-pareto}, we see that the sample average does not get close the the prior as time moves on. Instead, it becomes asymptotically parallel. These plots also compare with the well established Kaplan--Meier and Nelson--Aalen estimators. The simulation average approximates the true hazard and log-survival. 

We remark here that, if a user is not interested in computing certain statistics based on the sample paths of the posterior, it is also possible to compute the posterior mean via~\eqref{eq:posterior_mean_var} above, say, and use that as a proxy for whatever calculation one is interested in. The subtlety of such an approach is that there may be a bias. Indeed, for instance, taking the posterior mean of the hazard and then integrating it does \emph{not} result in the posterior mean of the log-survival, because the expectation does not commute with non-linear transformations. For this reason, it is generally better to compute the posterior mean of the sampled statistic of interest (say, via Monte Carlo) instead of the statistic of interest of the mean unless the computational complexity is prohibitive.

\subsection{Weibull-type synthetic data}\label{subsec:6-Weibull}

\begin{figure}[hbt!]
\centering{}
\includegraphics[width=6cm]{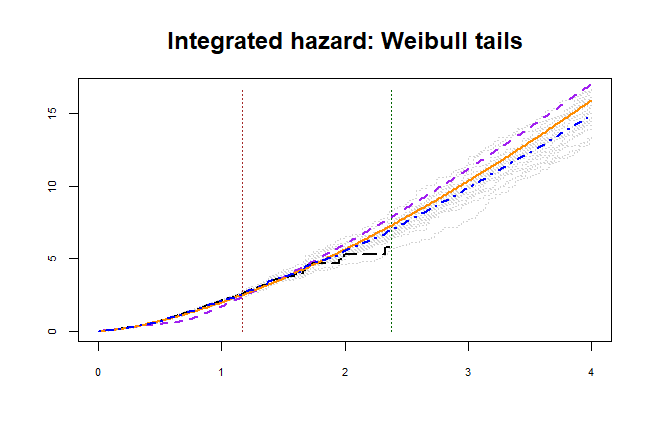}
\includegraphics[width=6cm]{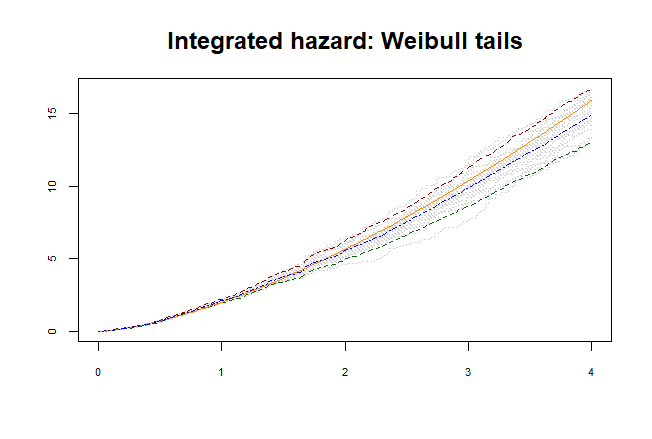}\\
\includegraphics[width=6cm]{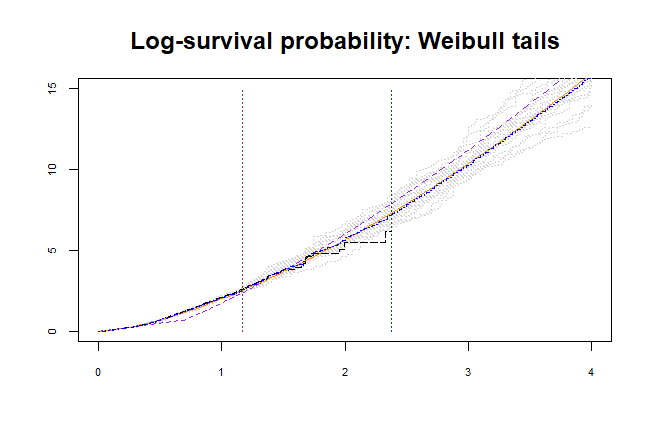}
\includegraphics[width=6cm]{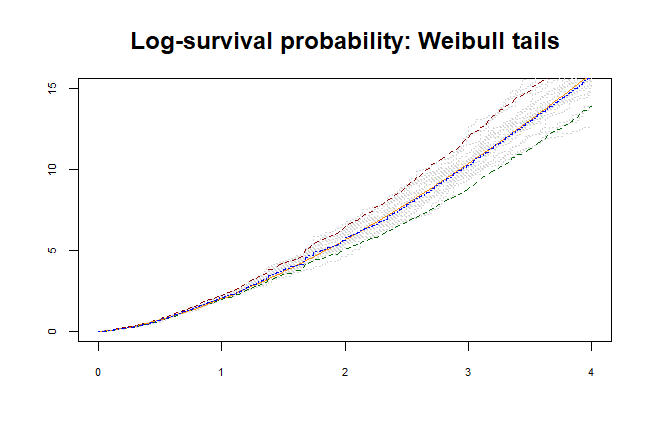}
\caption{Weibull tailed simulated data. We see 20 out of the 200 simulated paths dotted in light gray, the sample mean dot-dashed in blue, the prior dashed in purple, the Kaplan--Meier and Nelson--Aalen estimators long-dashed in black, the true hazard and log-survival solid in dark orange and, on the right, the upper and lower credible regions (generated via samples) long-dashed in red and green, respectively. The vertical dotted lines indicate where the $(n-k_n)$-th observation and where the largest observation are located.}
\label{fig:simulated-weibull}
\end{figure}

We consider simulated random variables $X$, generated via the rule  
\[
X=(E/\alpha)^{1 / (p + (1-E/\alpha)\vee 0)},
\]
where $E$ is a standard exponential random variable and $\alpha>0$ and $p>0$ are some parameters. It is easy to see that
\[
\pr(X>t)=\exp(-\alpha t^{p}),
\qquad\text{when }t\ge 1.
\]
We use the parameters $p=.5$ and $\alpha=2$ (i.e. the scale parameter presented in Subsection~\ref{subsec:weibull} above equals $l=(p/\alpha)^{1/p}=.0625$) and the same censoring mechanism of Subsection~\ref{subsec:6-Pareto}. Again, we draw $n=1000$ samples, assume from expert knowledge that our distribution is Weibull tailed (i.e., $p>0$) and use $a_n\coloneqq\log n$ and the quantile sequence $k_n\coloneqq \lceil2\sqrt{n}\rceil=64$ satisfying $k_n/n\to0$ and $k_n\to\infty$ to estimate $\hat p_n \approx 1.3722$ and $\hat\alpha_n\approx 3.6906$ (equivalently, $\hat l_n\approx .4862$). Despite the apparent bad fit, this estimation informs our baseline hazard function, given by:
\[
\frac{\dd}{\dd t}\Lambda_n(t)
= qI\{t<t_0\} 
    + \hat\alpha_n t^{\hat p_n-1}I\{t\ge 1\},
\quad t>0,
\]
where the parameters $q=1$ and $t_0=T_{n-k,n}$ are chosen as before. As in the Pareto case, in Figure~\ref{fig:simulated-weibull} we draw 200 samples of the posterior of both $H$ and $A$ and compare with the baseline functions, the true hazard and log-survival functions $-\log\ov F(t)$, as well as the corresponding Kaplan--Meier and Nelson--Aalen estimators. 

In Figure~\ref{fig:simulated-weibull}, we again see that the sample average becomes asymptotically parallel to the prior and also compares with the well with the Kaplan--Meier and Nelson--Aalen estimators in the range of observations, while staying very close to the true hazard past the last observation. In contrast to the Pareto example shown above, the estimation of the parameters is apparently not good. Despite this fact, the robustness of the least square methodology employed in Subsection~\ref{subsec:weibull} above makes the simulation average (which can also be observed of the posterior mean) a good approximation for the true integrated hazard and log-survival probabilities. 

The model's strength comes from breaking the fitting problem in two via a non-parametric splicing combining the strength of classical non-parametric models in the range where most events occur and the tail estimation (for instance, Pareto or Weibull tails). Our model and simulation method proved robust even with an apparent bad fit of the tails (and censoring mechanism of different tail behaviour to the true survival law) resulted in an accurate posterior mean.

%\subsection{Exact versus approximate simulations}

%\subsection{Non-parametric splicing and credible regions}

\subsection{Real data analysis}\label{sec:7}

We based our analysis on the Diabetic Retinopathy dataset, which comprises 394 eye-specific observations from 197 patients. This dataset, available in R (see also~\citealp{Huster_Brookmeyer_Self_1989}), stems from a $50\%$ random sample of patients identified as having “high-risk” diabetic retinopathy by the Diabetic Retinopathy Study (DRS). Each patient had one eye randomly assigned to receive laser treatment, while the contralateral eye served as an untreated control. The primary endpoint was the time from treatment initiation until visual acuity declined below 5/200 on two consecutive follow-up visits. Given that follow-up assessments were conducted every three months, this definition introduces a built-in lag of approximately six months. Consequently, the reported survival times represent the actual time to blindness (in years) adjusted by subtracting this minimum possible time to event (6.5 months). Censoring occurred due to patient death, study dropout, or administrative end of follow-up.

\begin{figure}[hbt!]
\centering{}
\includegraphics[width=6cm]{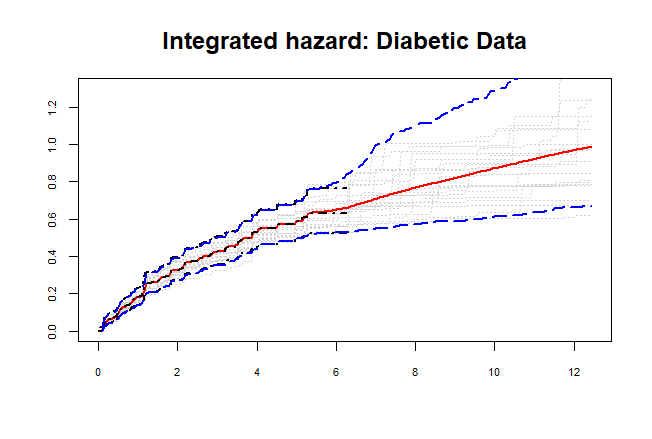}
\includegraphics[width=6cm]{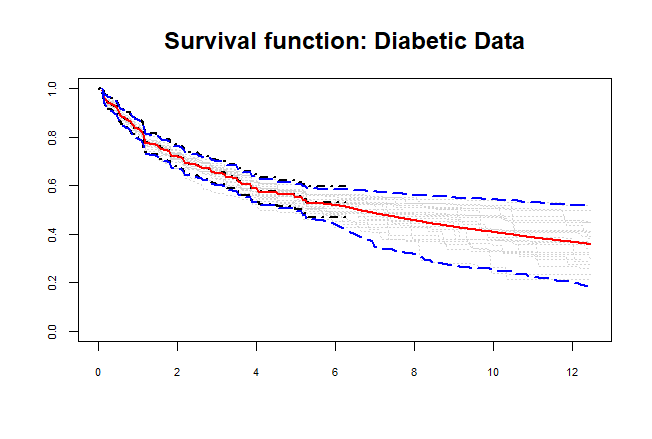}
\caption{Integrated hazard and survival functions for the diabetic data. The Kaplan--Meier (survival) and Nelson--Aalen (hazard) estimators (dotdashed in black) are depicted with their 95\% confidence intervals up to the last observation of the Diabetic dataset. The $x$-axis is in years. We also present the simulation average (solid line in red) of $1000$ sampled paths using our algorithms, along with the corresponding empirical 95\% credible regions (dashed in blue).}
\label{fig:diabetic}
\end{figure}

This data is known for having Weibull tails (which can be easily confirmed by performing a Weibull QQ-plot. We ran our methodology with this data, using $k_n\coloneqq\lceil2\sqrt{n}\rceil$, which resulted in the estimators $\hat p_n\approx 0.5144$, $\hat\alpha_n\approx .2138$ and $\hat l_n\approx 5.5112$. This informs our baseline hazard as in Subsection~\ref{subsec:6-Weibull} above, which gave a good fit as shown in Figure~\ref{fig:diabetic}. In particular, we remark the ability of the credible regions to extend beyond the observed data, thanks to extrapolation arising from the prior, which in turn is informed through extreme-value techniques.

The good match in Figure~\ref{fig:diabetic} of both sampled-based estimators with the Kaplan--Meier and Nelson--Aaeln estimators, as well as the good match of their corresponding credible regions up to the last observation (recall that splice occurs from the $(n-k_n)$-th observation), shows the robustness of our method. The main advantage of our model and simulation algorithms is that we are able to extend the estimator and their credible region well beyond the observed data.

\section{Conclusion and outlook}\label{sec:8}

This paper introduces a novel Bayesian non-parametric framework for survival analysis that addresses the limitations of traditional methods when dealing with time-to-event data. Our approach uses conditionally Lévy processes with stochastic hyperparameters, allowing for dynamic prior specification and adaptation to evolving data.  The key contributions are threefold:  first, the development of a flexible model that incorporates expert knowledge and adapts to data characteristics; second, the derivation of efficient and exact simulation algorithms for posterior inference, significantly improving upon existing approximate methods; and third, the application of the framework to construct non-parametric spliced models, which accurately capture both the body and tail behavior of survival distributions.  Our theoretical results establish Bayesian consistency and Bernstein-von Mises theorems, guaranteeing the validity of the model and its asymptotic agreement with frequentist and classical Bayesian counterparts.

The implications of our work are cross-disciplinary.  In medical research, for instance, the model can effectively analyze patient survival data exhibiting diverse patterns due to varying treatment responses or underlying health conditions. Similarly, in actuarial science, the model can accurately predict claim severities by accounting for both frequent, low-cost claims and rare, high-cost events.  The ability to incorporate external or internal information or expert judgment into the prior allows for extrapolation well beyond the observed data.

While our exact simulation algorithms represent a significant advance, their computational cost may be substantial for extremely large datasets.  Future research could focus on developing computationally efficient approximations that maintain accuracy.  Furthermore, the asymptotic analysis relies on specific regularity conditions on the Lévy measures; investigating the robustness of the model under relaxed or violated assumptions would be beneficial.  Exploring alternative prior specifications and hyperparameter choices is also a promising avenue for future work, including the investigation of other suitable tail families beyond Pareto and Weibull. Finally, extending the model to incorporate more sophisticated data structures, such as time-varying covariates or competing risks, warrants further investigation.

%%%%%%%%%%%%%%%%%%%%%%%%%%%%%%%%%%%%%%%%%%%%%%
%% Acknowledgements                         %%
%% should be provided in the                %%
%% Acknowledgements section.                %%
%%%%%%%%%%%%%%%%%%%%%%%%%%%%%%%%%%%%%%%%%%%%%%
\begin{acks}[Acknowledgments]
The authors would like to thank the Isaac Newton Institute for Mathematical Sciences, Cambridge, for support and hospitality during the programme Heavy tails in machine learning, where work on this paper was undertaken.
\end{acks}

%%%%%%%%%%%%%%%%%%%%%%%%%%%%%%%%%%%%%%%%%%%%%%
%% Funding information, if any,             %%
%% should be provided in the                %%
%% funding section.                         %%
%%%%%%%%%%%%%%%%%%%%%%%%%%%%%%%%%%%%%%%%%%%%%%
\begin{funding}
The first author (MB) was supported by the Carlsberg Foundation, grant CF23-1096.

The second author (JGC) was supported by DGAPA-PAPIIT grant 36-IA104425 and EPSRC grants EP/V009478/1 and EP/V521929/1.
\end{funding}

\begin{supplement}
\stitle{Proofs of the main theorems and the simulation of truncated gamma processes}
%\slink[url]{Link to the supplement}
\sdescription{The supplement provides the proof of Theorems~\ref{thm:consist} and~\ref{thm:BvM1}, as well as derive rule of thumb hyper-parameters to be used in the simulation of the truncated gamma process.}
\end{supplement}

\bibliographystyle{ba}
\bibliography{Bayesian_survival.bib}

\newpage 

\subsection*{Proof of Theorem~3.1}
As is standard, we show for $t>0$, $\E_{\nu_n^H,\mathcal{D}_n}[H(t)]\to H_0(t)$, and  $\mathrm{var}_{\nu_n^H,\mathcal{D}_n}[H(t)]\to0$, which by Pólya's theorem, implies uniform convergence on $[0,\tau]$.

By the mean equation (4) and the conjugacy result, Theorem~2.2, using the almost sure continuity, the posterior mean is given $\tilde{\mathcal{L}}_n-\mbox{a.s.}$ by
\begin{align*}
\int_{[0,t)}\int_0^1 (1-x)^{Y_n(s)}q_n(s,x)\dd x\Lambda_n(\dd s)
+\int_{[0,t)}\frac{\int_0^1x(1-x)^{Y_n(s)-1}q_n(s,x)\dd x}{\int_0^1(1-x)^{Y_n(s)-1}q_n(s,x)\dd x}\dd N_n(s)\\
=:I_1+I_2.
\end{align*}
where we have used that $\Delta N_n(t)=1$, whenever $N_n$ has a jump at $t$, by continuity of $F_0$. Let $c\in (0,c_2)$, and $c_2=\inf_{t\in[0,\tau]} \lim_n Y_n(t)/n=\inf_{t\in[0,\tau]} \bar{G}_0(t-)\bar{F}_0(t-)>0$. Then by the law of large numbers, and by the assumption $\kappa_n=o(n)$, $I_1$ is bounded for any $t<\tau$ for large enough $n$ by
\begin{align*}
\int_{[0,t)}\int_0^1 (1-x)^{Y_n(t)}q_n(s,x)\dd x\Lambda_n(\dd s)
\le Q_n\Lambda_n(\tau)\int_0^1 (1-x)^{nc}\dd x=\frac{\Lambda_n(\tau)Q_n(\tau)}{nc},
\end{align*}
which is vanishing by assumption~(8): $Q_n(\tau)\Lambda_n(\tau)=o(n)$ a.s., implying $I_1\to0$. Notice that $N_n/n$ is of bounded variation and almost surely, $\dd N_n/n \to \bar G_{0-}\dd F_0$. Thus we can restrict ourselves to analyze the terms inside of $I_2$. We may split the denominator in $I_2$ for any $\varepsilon$ as
\begin{align*}
\int_0^\varepsilon (1-x)^{Y_n(t)-1}q_n(t,x)\dd x
+\int_\varepsilon^1 (1-x)^{Y_n(t)-1}q_n(t,x)\dd x.
\end{align*}
Again, the assumptions imply that, for all sufficiently large $n$, the second term above is bounded by $(1-\varepsilon)^{cn}$. By letting $\varepsilon_n\to0$  
%slower than $n^{-a/(cn)}$, we get 
sufficiently slow in $(0,1)$, say 
\[
\varepsilon_n\sim -\log(1-\ve_n)=\omega(n^{-1}\log n),
\]
ensures that $(1-\varepsilon)^{cn}\le n^{-c\omega(1)}=o(n^{-a})$ for any $a>0$. Thus the integrand of $I_2$ is expanded as
\begin{align*}
&\frac{\int_0^{\varepsilon_n}x(1-x)^{Y_n(t)-1}q_n(t,x)\dd x+o(n^{-2})}{\int_0^{\varepsilon_n}(1-x)^{Y_n(t)-1}q_n(t,x)\dd x + o(n^{-2})}\\
&=
\frac{q_n^0(t)\{1+o(1)\}Y_n(t)^{-2}\int_0^{Y_n(t)\varepsilon_n}u(1-u/Y_n(t))^{Y_n(t)-1}\dd u+o(n^{-2})}{q_n^0(t)\{1+o(1)\}Y_n(t)^{-1}\int_0^{Y_n(t)\varepsilon_n}(1-u/Y_n(t))^{Y_n(t)-1}\dd u + o(n^{-2})}.
\end{align*}
For $\varepsilon_n$ going slowly to zero, the remaining integrals in the above expression converge $\tilde{\mathcal{L}}_\infty-\mbox{a.s.}$ to $\int_0^{\infty} ue^{-u}\dd u=1$ and $\int_0^{\infty} e^{-u}\dd u=1$. It follows that
\begin{align*}
I_1+I_2 \to0+ \int_{[0,t)}\frac{1}{\bar F_0(s-)\bar G_0(s-)} \bar G_0(s-)\dd F_0(s)=\int_{[0,t)}\frac{1}{\bar F_0(s-)}\dd F_0(s)=H_0(t).
\end{align*}
We now turn to the variance, which by (5) is given $\tilde{\mathcal{L}}_n-\mbox{a.s.}$ by
\begin{align*}
&\int_{[0,t)}\int_0^1 x(1-x)^{Y_n(s)}q_n(s,x)\dd x\Lambda_n(\dd s)+\\
&\int_{[0,t)}\Big[\frac{\int_0^1x^2(1-x)^{Y_n(s)-1}q_n(s,x)\dd x}{\int_0^1(1-x)^{Y_n(s)-1}q_n(s,x)\dd x}-\Big[\frac{\int_0^1x(1-x)^{Y_n(s)-1}q_n(s,x)\dd x}{\int_0^1(1-x)^{Y_n(s)-1}q_n(s,x)\dd x}\Big]^2\Big]\dd N_n(s)\\
&=J_1+J_2.
\end{align*}
Using similar arguments to the ones used for the posterior mean as well as the identity $\int_0^1 x(1-x)^p\dd x=[(p+1)(p+2)]^{-1}$ for $p\ge 0$, we can bound
\begin{align*}
J_1\lesssim \Lambda_n(\tau)Q_n(\tau)[(nc-\kappa_n+1)(nc-\kappa_n+2)]^{-1},
\end{align*}
which goes a.s. to zero by (8). Similarly, the first term inside the $J_2$ integral is equivalent to $2 Y_n(t)^{-2}$, while the second term inside $J_2$ is equivalent to $[Y_n(t)^{-1}]^2$, uniformly in $t\in[0,\tau]$. Both of them vanish when multiplied by $n$, which in turn implies $J_2\to0$. Thus the variance vanishes, completing the proof.

\subsection*{Proof of Theorem~3.3}
Part 1. We make use of the decomposition (3), whereby $H=H^{(1)}+H^{(2)}$ with $H^{(1)}$ a Poisson process and $H^{(2)}$ a fixed-jump process. From (4) and (5) and the conjugacy result, Theorem~2.2, we obtain
\begin{align*}
\E_{\nu_n^H,\mathcal{D}_n} H^{(1)}(t)&=\int_{[0,t)}\int_0^1 (1-x)^{Y_n(s)}q_n(s,x)\dd x \dd s\\
\mathrm{var}_{\nu_n^H,\mathcal{D}_n} H^{(1)}(t)&=\int_{[0,t)}\int_0^1 x(1-x)^{Y_n(s)}q_n(s,x)\dd x \dd s.
\end{align*}
Thus, by similar arguments as in the proof of Theorem~3.1, we get that 
\begin{align*}
\sqrt{n}\E_{\nu_n^H,\mathcal{D}_n} H^{(1)}(t) 
&%\lesssim \frac{\sqrt{n}}{nc-\kappa_n}
\lesssim Q_n(t) n^{-1/2}\\
\sqrt{n}\mathrm{var}_{\nu_n^H,\mathcal{D}_n} H^{(1)}(t) 
&%\lesssim \frac{\sqrt{n}}{n^2c^2-\kappa_n^2+nc-\kappa_n+2}
\lesssim Q_n(t)n^{-3/2},
\end{align*}
and both vanish by the assumption on the sequences $\kappa_n$ and $Q_n$. Both expressions are nondecreasing, and so pointwise convergence implies uniform convergence in $\mathbb{D}([0,\tau])$. Hence we concentrate on $H^{(2)}$, a process of finitely many jumps at event times. We apply Lyapunov's central limit theorem. By Theorem~2.2, the fourth moment of the summands of the process $H^{(2)}$ is seen to be given by
\begin{align*}
\E_{\nu_n^H,\mathcal{D}_n} \big[(\Delta H^{(2)}(t))^4\big]
&=\frac{\int_0^1x^4(1-x)^{Y_n(t)-1}q_n(t,x)\dd x}{\int_0^1(1-x)^{Y_n(t)-1}q_n(t,x)\dd x}.
\end{align*}
For $\varepsilon_n$ going slowly to zero technique from Theorem~3.1, the above expression is $\tilde{\mathcal{L}}_\infty-\mbox{a.s.}$ asymptotically equivalent to $Y_n(t)^{-4}\int_0^{\infty} s^4e^{-s}\dd s=24 \,Y_n(t)^{-4}$, which is of order $O_\pr(n^{-4})$. Thus the sum 
\begin{align*}
\sum_{s\le t} \sqrt{n}\big(\Delta H^{(2)}(s)|(\nu_n^H,\mathcal{D}_n)-\E_{\nu_n^H,\mathcal{D}_n} \Delta H^{(2)}(s)\big)
\end{align*}
satisfies Lyapunov's condition. We require calculating the variance, which corresponds to $J_2$ in Theorem~3.1 where the summands were seen to be of order $2Y_n(t)^{-2}-[Y_n(t)^{-1}]^2=Y_n(t)^{-2}$. Consequently
\begin{align*}
\sum_{s\le t} \mathrm{var}_{\nu_n^H,\mathcal{D}_n} \sqrt{n}\Delta H^{(2)}(s)
&=\int_{[0,t)}\frac{n}{Y_n(s)^2}\dd N_n(s)+o(1)\\
&\to\int_{[0,t)}\frac{\bar G_{0-}\dd F_0}{(\bar F_{0-}\bar G_{0-})^2}
=U_0(t).
\end{align*}
Thus, by Lyapunov's central limit theorem, marginal convergence is established to $B\circ U_0$. On $[0,\tau]$ the convergence of the variance is uniform and to a continuous limit. Increments of the approximating sequence and of the limit are independent, and so upon verification of condition (iii) of Theorem V.19 in \cite{pollard1984convergence}, we obtain the required convergence in the Skorokhod space $\mathbb{D}([0,\tau])$.

Part 2. Similar to the treatment of $I_2$ of Theorem~3.1, for $\varepsilon_n=\omega(n^{-1}\log n)$ as in the statement and $r_n(s,x)=q_n(s,x)/q_n^0(s)-1$ we get that
\begin{align*}
&\E_{\nu_n^H,\mathcal{D}_n} H-H_n=\\
&\int_{[0,\cdot)}\frac{\dd N_n(t)}{Y_n(t)}\Bigg[\frac{\int_0^{Y_n(t)\varepsilon_n}u(1-u/Y_n(t))^{Y_n(t)-1}(1+r_n(t,u/Y_n(t)))\dd u+o(n^{-3})}{\int_0^{Y_n(t)\varepsilon_n}(1-u/Y_n(t))^{Y_n(t)-1}(1+r_n(t,u/Y_n(t)))\dd u + o(n^{-3})}-1\Bigg].
\end{align*} 
Since $H_n$ is of bounded variation, it suffices to examine the term in square brackets. We obtain by replacing the upper (lower) bound above (below), that the square bracket term is bounded above, $\pr-$a.s. for large $n$ by
\begin{align*}
\frac{\Gamma(2)+C_n\Gamma(2+\alpha)\frac{1}{Y_n(t)^\alpha}+O_\pr(1/Y_n(t)^{\alpha})+o(n^{-3})}{\Gamma(1)-C_n\Gamma(1+\alpha)\frac{1}{Y_n(t)^\alpha}+O_\pr(1/Y_n(t)^{\alpha})+o(n^{-3})}-1\\
=O_\pr([1\vee C_n]/Y_n(t)^{\alpha}),
%=O_\pr(n^{0\vee \eta-\alpha}),
\end{align*} 
uniformly in $t$. Multiplying by $\sqrt{n}$ now provides the claim of Part 2. Part 3 follows from Parts 1 and 2.

\subsection*{Simulation of the truncated gamma process}

Given the high sensitivity of $C(\vartheta,\delta,\mu)$ in Subsection~4.3 in $\zeta$ (and hence $\vartheta$), our first goal is to approximately minimise the second term as a function of $\vartheta>0$. First note that $\mu\mapsto\zeta$ is bounded as $\mu\to 0$, at which point we may take $\vartheta=1$ and $\delta=1/2$, so our main concern is the appropriate selection of parameters for large values of $\mu$.

First use Stirling's approximation to see that $(\vartheta,\delta)\mapsto C(\vartheta,\delta,\mu)$ is approximately proportional to the exponential of
\[
\zeta e^\zeta+e^\zeta+\delta-(e^\zeta+\delta-1/2)\log(e^\zeta+\delta)-\log(\vartheta\delta(1-\delta)).
\]
If $\mu$ is large, then $\log(e^\zeta+\delta)$ is approximately equal to $\zeta$, so the display above is approximately equal to
\[
e^\zeta-(\zeta-1)(\delta-1/2)
-\log(\vartheta\delta(1-\delta)) + 1/2.
\]
Clearly, $\vartheta$ should not be chosen larger than $1$ in general, so $\zeta\sim\log\mu$ as $\mu\to\infty$.

Optimising over $\delta$ leads to the equation
\[
\frac{1}{1-\delta}=\zeta-1+\frac{1}{\delta}
\implies
(\zeta-1)\delta^2 + (3-\zeta)\delta - 1 = 0,
%(\zeta-1)\delta(1-\delta) + 1 - 2\delta = 0
\]
whose only solution on $(0,1)$ (for large $\mu$ and hence $\zeta'=\zeta-1>2$) is 
\[
\delta 
= \frac{\zeta-3+\sqrt{(\zeta-1)^2+4}}{2(\zeta-1)}
=\frac{1}{2}-\frac{1}{\zeta-1}
+\frac{1}{2}\sqrt{1+4(\zeta-1)^{-2}},
\]
which implies that $1-\delta\sim (\zeta-1)^{-1}-(\zeta-1)^{-2}$ as $\zeta\sim \log\mu\to\infty$. Since $\vartheta$ is not chosen larger than $1$, for simplicity (which ensures $\delta=1/2$ for $\mu=0$), we pick $\delta=1-\log(e^2+\mu)^{-1}$. Then, the main expression to control is $e^{\log\mu+\vartheta}-\log\vartheta$, which leads to $\vartheta \sim 1/\mu$ as $\mu\to\infty$, motivating the choice $\vartheta=(1+\mu)^{-1}$. 

%%%%%%%%%%%%%%%%%%%%%%%%%%%%%%%%%%%%%%%%%%%%%%
%% Supplementary Material, including data   %%
%% sets and code, should be provided in     %%
%% {supplement} environment with title      %%
%% and short description. It cannot be      %%
%% available exclusively as external link.  %%
%% All Supplementary Material must be       %%
%% available to the reader on Project       %%
%% Euclid with the published article.       %%
%%%%%%%%%%%%%%%%%%%%%%%%%%%%%%%%%%%%%%%%%%%%%%
%\begin{supplement}
%\stitle{Title of Supplement A}
%\sdescription{Short description of Supplement A.}

\end{document}